\documentclass[11pt,a4paper]{article}
\usepackage{url}
\usepackage{jcappub}

\usepackage{graphicx}
\usepackage{dcolumn}
\usepackage{bm}
\usepackage{amsfonts}
\usepackage{color}
\usepackage{newtxtext,amsmath}
\usepackage{appendix}
\usepackage{subfigure}

\usepackage{lineno}
\definecolor{brightpink}{rgb}{1.0, 0.0, 0.5}

\usepackage{tabu}
\usepackage{booktabs}
\usepackage{tabularx}
\usepackage{lineno}

\providecommand{\aap}{Astron.\ Astrophys.}
\providecommand{\apj}{Astrophys. J.}
\providecommand{\apjs}{Astrophys.\ J.\ Suppl.}
\providecommand{\apjl}{Astrophys.\ J.}
\providecommand{\jcap}{JCAP}
\providecommand{\mnras}{MNRAS}
\providecommand{\nphysa}{Nucl.\ Phys.\ A}
\providecommand{\pasa}{Publications of the Astronomical society of Australia}
\providecommand{\physrep}{Phys.\ Rep.}
\providecommand{\prl}{Phys.\ Rev.\ Lett.}
\providecommand{\prd}{Phys.\ Rev.\ D}
\providecommand{\prc}{Phys.\ Rev.\ C}
\providecommand{\nat}{Nature}

\begin{document}

\title{Hyperons during proto-neutron star deleptonization and the emission of dark flavoured particles}

\author[a]{Tobias~Fischer,}
\author[b,c,d]{Jorge~Martin~Camalich,}
\author[e,f]{Hristijan~Kochankovski,}
\author[g,h,i]{Laura~Tolos}

\affiliation[a]{\em Institute for Theoretical Physics, University of Wroclaw, \\ Pl. M. Borna 9, 50-204 Wroclaw, Poland}
\affiliation[b]{\em Instituto de Astrof\'isica de Canarias, \\ C/ V\'ia L\'actea, s/n E38205 - La Laguna, Tenerife, Spain}
\affiliation[c]{\em Departamento de Astrof\'isica, Universidad de La Laguna, \\ La Laguna, Tenerife, Spain}
\affiliation[d]{\em CERN, Theoretical Physics Department, \\ CH-1211 Geneva 23, Switzerland}
\affiliation[e]{\em Departament de F\'{\i}sica Qu\`antica i Astrof\'{\i}sica and Institut de Ci\`encies del Cosmos, \\ Universitat de Barcelona, \\ Mart\'i i Franqu\`es 1, 08028, Barcelona, Spain}
\affiliation[f]{\em Faculty of Natural Sciences and Mathematics-Skopje, \\ Ss. Cyril and Methodius University in Skopje, \\ Arhimedova, 1000 Skopje, North Macedonia}
\affiliation[g]{\em Institute of Space Sciences (ICE, CSIC), \\ Campus UAB,  Carrer de Can Magrans, 08193 Barcelona, Spain}
\affiliation[h]{\em Institut d'Estudis Espacials de Catalunya (IEEC), \\ 08860 Castelldefels (Barcelona), Spain}
\affiliation[i]{\em Frankfurt Institute for Advanced Studies,  \\ Ruth-Moufang-Str. 1, 60438 Frankfurt am Main, Germany}
\emailAdd{tobias.fischer@uwr.edu.pl, jcamalich@iac.es, hriskoch@fqa.ub.edu, tolos@ice.csic.es}

\abstract{
Complementary to high-energy experimental efforts, indirect astrophysical searches of particles 
beyond the standard model have long been pursued.
The present article follows the latter approach and considers, for the first time, the self-consistent 
treatment of the energy losses from dark flavoured particles produced in the decay of hyperons 
during a core-collapse supernova (CCSN). 
To this end, general relativistic supernova simulations in spherical symmetry are performed, 
featuring six-species Boltzmann neutrino transport, and covering the long-term evolution of the 
nascent remnant proto-neutron star (PNS) deleptonization for several tens of seconds. 
A well-calibrated hyperon equation of state (EOS) is therefore implemented into the supernova 
simulations and tested against the corresponding nucleonic model. 
It is found that supernova observables, such as the neutrino signal, are robustly insensitive to 
the appearance of hyperons for the simulation times considered in the present study.
The presence of hyperons enables an additional channel for the appearance of dark sector 
particles, which is considered at the level of the $\Lambda$ hyperon decay.
Assuming massless particles that escape the PNS after being produced, these channels 
expedite the deleptonizing PNS and the cooling behaviour. 
This, in turn, shortens the neutrino emission timescale. 
The present study confirms the previously estimated upper limits on the corresponding 
branching ratios for low and high mass PNS, by effectively reducing the neutrino emission 
timescale by a factor of two. 
This is consistent with the classical argument deduced from the neutrino detection associated with SN1987A. 
}


\maketitle

\section{Introduction}
\label{sec:intro}
Massive stars end their lifes in the event of a CCSN, when the stellar core collapses due to pressure losses from electron captures on protons bound in iron-group nuclei and the photodissociation of nuclei. The core collapse halts when the density exceeds the normal nuclear density, and the stellar core bounces back with the formation of a shock wave. The initial propagation of this bounce shock leads to the release of a $\nu_e$-burst when reaching the neutrinospheres of the last inelastic scattering. The associated energy loss, in combination with the energy loss due to the photodissociation of the still infalling matter onto the expanding dynamic bounce shock, causes the latter to stall and turn into an accretion shock. Revival of this standing accretion shock is considered the CCSN explosion engine (for recent reviews, c.f., Refs.~\cite{Janka07,mirizzi16,Burrows2021Natur589} and references therein). 

Most of the gravitational binding energy gain is stored in the central compact PNS, in the magnetic-rotational energy~\cite{LeBlancWilson70}, as well as in thermal degrees of freedom and neutrinos of all flavours, on the order of several $10^{53}$~erg. Hence, the neutrino-heating mechanism for shock revival, first proposed in ref.~\cite{Bethe85}, has long been studied in modern CCSN simulations. These are based in multi-dimensional neutrino radiation hydrodynamics featuring neutrino transport, where in the presence of multi-dimensional hydrodynamics phenomena, such as convection and hydrodynamic instabilities, the neutrino-heating efficiency increases, compared to spherically symmetric CCSN models, which therefore generally fail to yield CCSN explosions. There are two exceptions, {\em (i)} low-mass stellar progenitors in the zero-age main sequence (ZAMS) mass range of 8--10~M$_\odot$ \cite{kitaura,Fischer10,Huedepohl10} as well as ultra-stripped stellar progenitors that result from binary evolution \cite{Tauris15,De:2018Sci362}, despite substantial differences between spherically symmetric and multi-dimensional simulations \cite{BMuller18,BMuller19,Ertl:2020ApJ890,Stockinger:2020MNRAS496,Wang24}, and {\em (ii)} CCSN explosions triggered due to a sufficiently strong first-order phase transition from normal nuclear matter to the quark-gluon plasma at high baryon density (c.f. Refs.~\cite{Sagert09,Fischer18,Zha20,Fischer:2021,Kuroda2022,KhosraviLargani2024ApJ964} and references therein) as well as the spontaneous scalarization of beyond the standard model degrees of freedom \cite{KurodaShibata:2023PhRvD107}.  

The latter CCSN explosion mechanism relates to one of the largest uncertainties in CCSN modelling, namely, the state of matter under CCSN conditions, i.e. at high baryon density, in excess of saturation density, large isospin asymmetry given by the hadronic charge density  (the proton abundance in the absence of other charged hadronic resonances, on the order of $Y_p=0.05$--$0.6$) and temperatures on the order of several tens of MeV 
(for a recent review of the role of the EOS in simulations of CCSN, see ref.~\cite{Fischer17}). 
Besides the QCD deconfinement phase transition, which has lately been explored extensively also in the context of binary neutron star mergers \cite{Bauswein19,Most19}, the impact of hadronic degrees of freedom has been studied in the context of failed CCSN in Refs.~\cite{Sumiyoshi2009ApJ690,Fischer09,O'Connor11,Nakazato2012ApJ745,Bandyopadhyay2015ApJ809}, even though the potential impact of strange, as well as non-strange hadronic  degrees of freedom has long been studied for neutron stars (see recent reviews \cite{FiorellaBurgio2018NuclearSupernovae,Tolos:2020aln,Burgio:2021vgk,MUSES:2023hyz} and references therein.) 
Thereby, particular emphasis has been placed on the {\em hyperon puzzle}, which is related to the softening of the supersaturation density EOS due to the presence of additional, heavy hadronic degrees of freedom, potentially violating the maximum neutron star mass constraint of about $2$~M$_\odot$, the latter was derived from observations of massive pulsars \cite{Demorest2010ShapiroStar,Antoniadis13,Fonseca:2021,Cromartie2020RelativisticPulsar,Fonseca:2021,Romani:2022jhd}.
Several possible solutions have been widely discussed in the literature. Among them, it has been put forward an early onset of the quark-hadron deconfinement phase transition at densities below which strange hadronic resonances would appear, or the existence of much more repulsive hadronic interactions at high baryon densities. 
However, hyperon-nucleon and hyperon-hyperon interactions are less understood than nucleon-nucleon interactions, due to the sparse knowledge of scattering data. This makes their potentials poorly constrained and dynamics far from understood (c.f. ref.~\cite{Petschauer:2020urh} and references therein). 
The theoretical understanding has been brought forward to solve the problem of thermal production of hyperons at low baryon density for heavy-ion physics based on the coupled channel approach (c.f. Refs.~\cite{Pok:2020PhRvD102,Pok:2021PhRvC103,Pok:2021EPJA57} and references therein) and furthermore, final-state interaction analyses \cite{CLAS:2021gur,J-PARCE40:2021qxa,J-PARCE40:2021bgw,J-PARCE40:2022nvq} as well as femtoscopy studies (see ref.~\cite{Fabbietti:2020bfg} for a review and references therein) have become available as well as data on hypernuclear structure \cite{Feliciello:2015dua,Tamura:2013lwa,Gal:2016boi}. 

Due to the poorly known hyperon EOS at high baryon density, effective hyperonic model EOS have long been employed in astrophysical studies, e.g., based on the relativistic mean-field (RMF)  approach that treats the unknown interactions via baryon-meson couplings. The present article implements the FSU2H$^*$ hyperonic relativistic mean field model in simulations of CCSN, focusing on the PNS deleptonization phase, i.e. the evolution after the CCSN explosion onset has been triggered during which the nascent PNS deleptonizes via the emission of neutrinos of all flavours on a timescale of 10~seconds. This phase is ideal for studying the impact of hadronic physics as the central density increases continuously as a direct consequence of the deleptonization. 

The requirements for multi-purpose EOS for applications of CCSN studies is to cover a large parameter space. 
Therefore, the RMF framework has long been developed (for recent reviews, see Refs.~\cite{Raduta:2021coc,Raduta:2022elz} and references therein). 
Recent reviews of the RMF framework for hyperon EOS can be fond in ref.~\cite{Oertel:2016bki,Burgio:2021vgk}\footnote{Note the existence of the online service CompOSE repository that provides data tables for different state of the art EOS ready for use in astrophysical applications, nuclear physics and beyond \cite{Dexheimer:2022qhn,CompOSECoreTeam:2022ddl}.}.

In addition to the impact that hyperons can potentially have in CCSN through the EOS, they can open potentially new cooling channels with the emission of novel, yet hypothetical, particles from the interior of the PNS. 
In particular, if the net energy carried away by these particles is comparable to that of neutrinos, then significant changes in the neutrino signal are predicted. 
These can be tested using the neutrino flux that was observed from SN1987A ~\cite{Raffelt:1987yt,Raffelt:1996wa}. Such phenomenology has been studied extensively in the context of the QCD axions, produced in the stellar plasma by, e.g. nucleon-nucleon bremsstrahlung \cite{Turner:1987by,Mayle:1987as,Burrows:1988ah,Burrows:1990pk,Carenza:2019pxu,Choi:2021ign,DiLuzio:2020wdo}, from pions studied \cite{Carenza:2020cis} and in simulations \cite{Fischer:2021jfm}, and has been extended to many other ``dark-sector'' theories predicting the existence of new weakly interacting particles with mass $\lesssim \mathcal O(T_{\rm PNS})$ \cite{Rrapaj:2015wgs,Chang:2018rso,Calore:2021klc,Balaji:2022noj,Lella:2022uwi,Manzari:2023gkt}, which is on the order of few tens of MeV. 

Adding ``standard'' degrees of freedom in the plasma, on top of nucleons and electrons/positrons, can also extend these analyses, as was recently demonstrated in the case of muons (c.f. Refs.~\cite{Bollig:2020xdr,Calibbi:2020jvd,Croon:2020lrf, Caputo:2021rux,Manzari:2023gkt} and references therein),
for pions \cite{Turner:1991ax,Raffelt:1993ix,Keil:1996ju,Carenza:2020cis,Fischer:2021jfm,Shin:2022ulh}, 
and for hyperons \cite{MartinCamalich:2020dfe,Camalich:2020wac,Alonso-Alvarez:2021oaj,Cavan-Piton:2024ayu}. 
In the case of hyperons, this was studied for various models that can induce flavour-changing neutral currents, 
such as the following decays, $\Lambda\to n$ and $\Lambda\to\gamma$, where the energy difference between final and initial states is considered to be carried away by dark degrees of freedom. 
Therefore, results of spherically symmetric simulations were employed without exotic cooling and implementing conventional EOS without hyperons. To establish limits on these models, the impact of the new dark emission on the dynamics of the CCSN was neglected, and thermodynamic quantities such as chemical potentials were re-derived by using interpolation tables of the hyperonic extensions of the EOS used in the simulations 
(see also ref.~\cite{Cavan-Piton:2024ayu} for a slightly different approach).

A dark sector coupled to quarks can have a rich flavour structure that is subject to strangeness-changing neutral currents (for reviews, see Refs.~\cite{Kamenik:2011vy,Goudzovski:2022vbt}). 
One prototypical example of these dark flavoured sectors is the QCD axion, with non-diagonal flavour couplings, known as ``familon''. 
This was first predicted in refs.~\cite{Davidson:1981zd,Wilczek:1982rv} as a consequence of jointly solving the strong CP and flavour problems, and was further developed \cite{Feng:1997tn,Calibbi:2016hwq,Ema:2016ops,MartinCamalich:2020dfe,DiLuzio:2023ndz}.
Other examples include the dark photon, which interacts with standard-model fermions only through non-renormalizable operators 
(see Refs.~\cite{Holdom:1985ag,Dobrescu:2004wz,Gabrielli:2016cut,Eguren:2024oov} and references therein) or dark baryons \citep{Alonso-Alvarez:2021oaj}. 
The latter are neutral fermions with masses of $\approx$1~GeV endowed with baryon number and that can be part of dark matter, and explain baryogenesis 
(c.f. Refs.~\cite{Elor:2018twp,Bringmann:2018sbs}) and the neutron lifetime puzzle \cite{Fornal:2018eol}.
The self-consistent implementation of hyperons in CCSN simulations enables us to overcome previous limitations 
(c.f. ref.~\cite{Camalich:2020wac} and references therein), studying the feedback of dark boson production on the supernova dynamics and neutrino emission.  

The paper is organized as follows.
In section~\ref{sec:RMF} the RMF framework for reference nucleonic and hyperonic EOS will be introduced, 
which is evaluated at selected conditions in section~\ref{sec:eos}, including finite temperature, and discussed accordingly. 
CCSN simulations will be launched and evaluated in section~\ref{sec:PNSdelept}. 
The comprehensive analysis from the impact of dark bosons will be provided in section~\ref{sec:DM} 
and the manuscript closes with a summary in section~\ref{sec:summary}.

\section{Hyperons within the relativistic mean field framework}
\label{sec:RMF}
The present study of the impact of strange degrees of freedom in CCSN simulations employed the FSU2H$^*$ RMF EOS. In this framework, the interactions between the baryons are mediated via the exchange of virtual mesons, based on the following Lagrangian for baryons $\mathcal{L}_{\rm B}$ and mesons $\mathcal{L}_{\rm m}$,
\begin{eqnarray}
{\cal L} &=& \sum_{\rm B} {\cal L}_{\rm B} + {\cal L}_{\rm m}~,
\end{eqnarray}
with
\begin{eqnarray}
{\cal L}_{\rm B} 
&=& 
\bar{\Psi}_{\rm B}
(
i\gamma_{\mu}\partial^{\mu} 
- q_{\rm B}{\gamma}_{\mu} A^{\mu} - m_{\rm B} 
+ g_{\sigma \rm B}\sigma + g_{\sigma^{*} \rm B}  \sigma^{*}
\nonumber \\
&-& 
g_{\omega \rm B}\gamma_{\mu} \omega^{\mu} 
- g_{\phi \rm B}\gamma_{\mu} \phi^{\mu} 
- g_{\rho \rm B}\gamma_{\mu} \, \vec{I}_{\rm B} \cdot \vec{\rho\,}^{\mu}
)
\Psi_{\rm B}~, 
\end{eqnarray}
and
\begin{eqnarray}
{\cal L}_{\rm m} &=& \frac{1}{2}\partial_{\mu}\sigma \partial^{\mu}\sigma - \frac{1}{2}m^2_{\sigma}\sigma^2 - \frac{\kappa}{3!}(g_{\sigma N}\sigma)^3 - \frac{\lambda}{4!}(g_{\sigma N}\sigma)^4 
 \nonumber \\ 
&+& \frac{1}{2}\partial_{\mu}\sigma^{*} \partial^{\mu}\sigma^{*}  
-\frac{1}{2}m^2_{\sigma^{*}}{\sigma^{*}}^2 \nonumber \\
&-&\frac{1}{4}\Omega^{\mu \nu}\Omega_{\mu \nu}  
+\frac{1}{2}m^2_{\omega} \omega_{\mu} {\omega}^{\mu} +  \frac{\zeta}{4!} g_{\omega N}^4 (\omega_{\mu}\omega^{\mu})^2 \nonumber \\
&-&\frac{1}{4}\vec{R}^{\mu \nu}\vec{R}_{\mu \nu} + \frac{1}{2}m^2_{\rho}\vec{\rho}_{\mu} \textcolor{blue}{\cdot} \vec{\rho\,}^{\mu}+ 
\Lambda_{\omega}g^2_{\rho N}\vec{\rho_{\mu}}\vec{\rho\,}^{\mu} g^2_{\omega N} \omega_{\mu} \omega^{\mu} \nonumber \\
&-& \frac{1}{4}P^{\mu \nu}P_{\mu \nu}
+\frac{1}{2}m^2_{\phi}\phi_{\mu}\phi^{\mu}-\frac{1}{4}F^{\mu \nu}F_{\mu \nu}~.
\label{eq:lagrangian}
\end{eqnarray}
The quantity $\Psi_{\rm B}$ represents the baryon Dirac field and $m_i$ indicates the mass of particle $i$, while the mesons that mediate the interaction between the baryons are two isoscalar-scalar mesons ($\sigma$ and $\sigma^{*}$), two isoscalar-vector mesons ($\omega$ and $\phi$), and one isovector-vector meson ($\rho$). The mesonic strength tensors are labeled with $\Omega_{\mu \nu} = \partial_{\mu} \omega_{\nu} -\partial_{\nu} \omega_{\mu} $, $\vec{R}_{\mu \nu} = \partial_{\mu} \vec{\rho_{\nu}} - \partial_{\nu} \vec{\rho_{\mu}}$, $P_{\mu \nu} = \partial_{\mu} \phi_{\nu} -\partial_{\nu} \phi_{\mu} $ and $F_{\mu \nu} = \partial_{\mu} A_{\nu} -\partial_{\nu} A_{\mu}$, whereas $\gamma^{\mu}$ are the Dirac matrices, $g_{\rm m \rm B}$ labels the coupling of baryon ${\rm B}$ to meson ${\rm m}$ and $\vec{I}_{\rm B}$ is the isospin operator. We stress that, since we are dealing with charge-neutral objects in the absence of magnetic fields, the electromagnetic part of the Lagrangian does not play a role. 

\begin{table*}
\begin{center}
\caption{Parameters of the model FSU2H$^*$, meson masses $m_i$ and meson-nucleon coupling constants $g_i^2$. The mass of the nucleon is equal to $m_N= 939$~MeV.}
\begin{tabular}{|ccccccccccccccc|}
\hline
$m_{\sigma}$ & & $m_{\omega}$ & & $m_{\rho}$ & & $m_{\sigma^{*}}$ & & $m_{\phi}$ & & $g_{\sigma N}^2$ & & $g_{\omega N}^2$ & & $g_{\rho N}^2$ \\
$[$MeV$]$ & & $[$MeV$]$ & & $[$MeV$]$ & & $[$MeV$]$ & & $[$MeV$]$ & & & & & & \\
\hline
497.479 & & 782.5 & & 763 & & 980 & & 1020 & & 102.72 & & 169.53 & & 197.27 \\
\hline
\end{tabular}
\label{table:nuclearparam}
\end{center}
\end{table*}

\begin{table*}
\begin{center}
\caption{Other parameters of the Lagrangian~\eqref{eq:lagrangian} of the model FSU2H$^*$.}
\begin{tabular}{|cccccccccc|}
\hline
$\kappa$ & && $\lambda$ & && $\zeta$ & && $\Lambda_{\omega}$ \\
$[$MeV$]$ & && & && & &&\\
\hline
4.0014 & && -0.0133 & && 0.008 & && 0.045 \\
\hline
\end{tabular}
\label{table:nuclearparam_2}
\end{center}
\end{table*}

In order to determine the EOS of the system, one first needs to obtain the Euler-Lagrange equations of motion for each particle. Then, a closed set of equations can be obtained by using the mean-field approximation that allows for the meson fields to be replaced by their expectation values, as well as by coupling those equations with the weak interaction equilibrium condition and baryon and charge conservation number equations. The solution of this set fixes the composition of matter and the values of the meson fields. Finally, from the stress-energy tensor all thermodynamic quantities can be obtained, such as the pressure and energy density (for details on the set of equations to be solved, we refer the reader to the Refs.~\cite{Tolos:2016hhl,Tolos:2017lgv,Kochankovski:2022rid,Kochankovski2024MNRAS528} and references therein).

We show in the following the explicit form of the effective mass and chemical potential of the different species, since these are important quantities whose behaviour influences the thermal evolution in CCSN explosions. The effective mass of a given particle reads as follows,
\begin{equation}
m_{\rm B}^{*} = m_{\rm B} - g_{{\rm B} \sigma}  \bar{\sigma} - g_{{\rm B} \sigma^*} \bar{\sigma}^{*}~,
\label{eq:meff}
\end{equation}
where the expectation values of the scalar mesons, denoted as $\bar{\sigma}$ and $\bar{\sigma}^{*}$, govern their behaviour.
Since no isospin splitting is considered due to the absence of the $\delta$ meson channel, $m_N=939$~MeV is used in FSU2H$^*$ for the nucleon rest masses while those of the hyperons are given in the Particle Data Booklet~\cite{PDG}. The expectation values
of the vector mesons, i.e. $\bar\omega$, $\bar\rho$ and $\bar\phi$, control the different chemical potentials as follows,
\begin{equation}
\mu_{\rm B}^{*} = \mu_{\rm B} - g_{{\rm B}\omega}\bar \omega  - g_{{\rm B}\rho} I_{3{\rm B}} \bar \rho- g_{{\rm B}\phi}\bar \phi~,
\label{eq:mueff}
\end{equation}
with the thermodynamic chemical potential $\mu_{\rm B}$. We note that the hidden mesons $\sigma^{*}$ and $\phi$ are coupled only to particles with non-zero strangeness.

\section{Equation of state with hyperons}
\label{sec:eos}
The different parameters of the FSU2H$^*$ model are chosen so as to predict an EOS that is compatible with the constraints coming from heavy ion collisions as well as those obtained from the saturation properties of nuclear matter and finite nuclei. 
At the same time the model predicts a maximum neutron star mass that is above 2.0~M$_{\odot}$ and a radius  about 13~km for canonical neutron stars of 1.4~M$_\odot$ (for details, see Refs.~\cite{ Tolos:2016hhl,Tolos:2017lgv}). 
This model was then extended to finite temperature \cite{Kochankovski:2022rid}, whereas more recently the hyperonic uncertainties have been analyzed for the determination of the masses, radii, tidal deformabilities and moments of inertia, both at zero and finite temperatures \cite{Kochankovski2024MNRAS528}. 
Indeed, the FSU2H$^*$ model is compatible with the tidal deformability extracted from the GW170817 event \cite{LIGOScientific:2017vwq} as well as the NICER determinations on radii \cite{NICER_Miller2019,NICER_Watts2019,NICER_Miller2021,NICER_Riley2021}. 
The values of the parameter meson masses and meson-nucleon coupling constants can be found in Table~\ref{table:nuclearparam}, 
whereas the other parameters of the Lagrangian~\eqref{eq:lagrangian} are given in Table~\ref{table:nuclearparam_2}. The ratios of coupling constants of hyperons to mesons are give in Table~\ref{table:hypparam}. 

In the present work we make use of the FSU2H$^*$ model in two different ways. 
Firstly, a set of simulations is performed with the exact FSU2H$^*$ EOS, and the results of those simulations 
are referred as HYPERONS. 
The results of the second set of simulations, that will be referred as NUCLEONS, are obtained within the 
same FSU2H$^*$ model, but preventing the appearance of hyperons  by setting their masses to infinity. 
This set is necessary in order to quantify the impact that hyperons have on the observables. 

\begin{table}
\begin{center}
\caption{Ratios of coupling constants of hyperons to mesons with respect to the nucleonic ones.}
\begin{tabular}{|llllllllllllllll|}
\hline
$Y$ & && $R_{\sigma Y}$ & && $R_{\omega Y} $ & && $R_{\rho Y}$ & && $R_{\sigma^{*}Y}$ & && $R_{\phi Y}$ \\
\hline
$\Lambda$ & && $0.6113$ & && $2/3$ & && $0$ & && $0.2812$ & && $-\sqrt{2}/3$  \\
$\Sigma$ & && $0.4673$ & && $2/3$ & && $1$ & && $0.2812$ & && $-\sqrt{2}/3$  \\
$\Xi$ & && $0.3305$ & && $1/3$ & && $1$ & && $0.5624$ & && $-2\sqrt{2}/3$  \\
\hline
\end{tabular}
\end{center}
\label{table:hypparam}
\end{table}

\begin{figure*}[htp]
\begin{center}
\includegraphics[angle=0.,width=1\columnwidth]{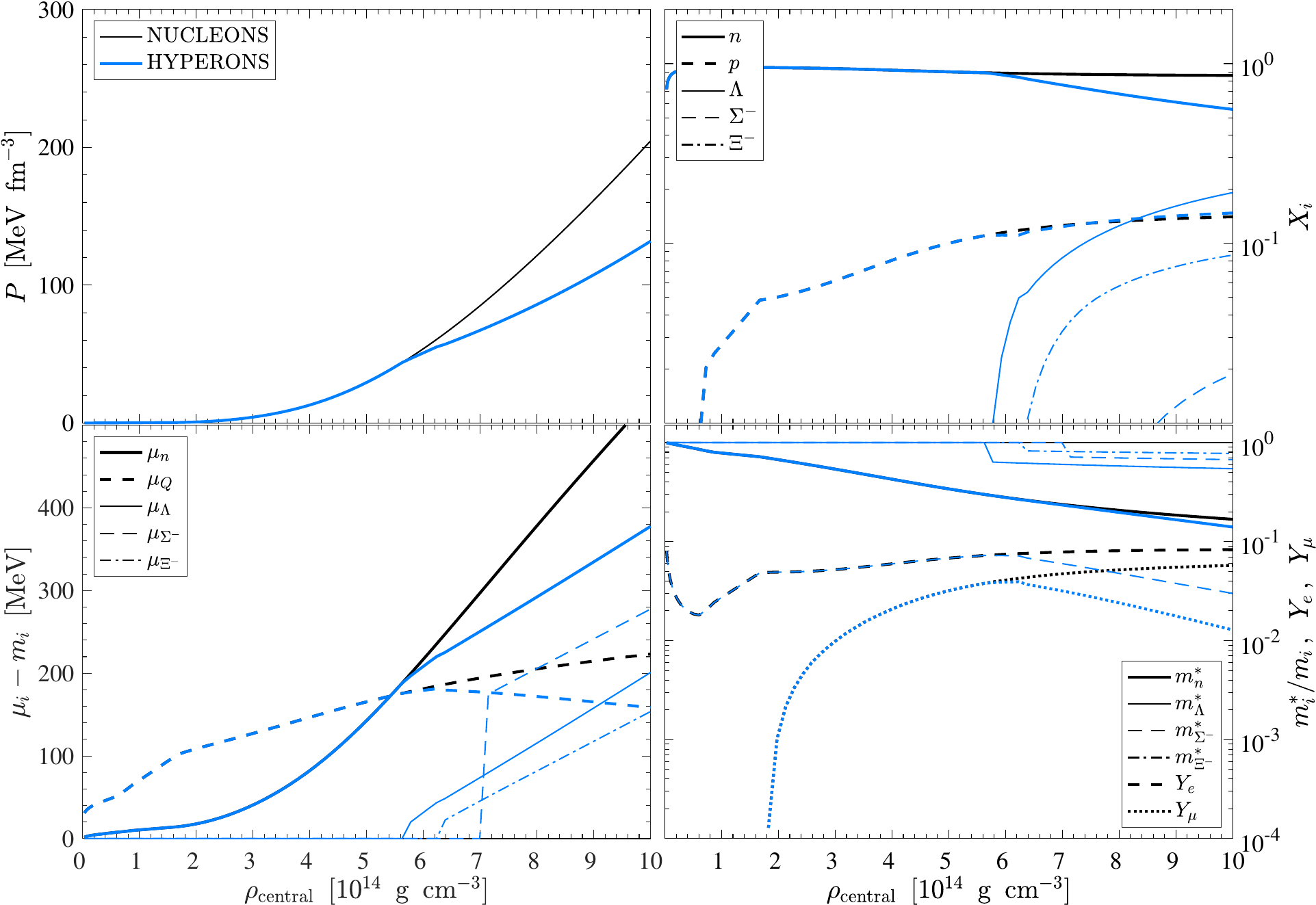}
\caption{~Equation of state at $\beta$-equilibrium at $T=0$, comparing the HYPERON EOS(blue lines) and
 the reference NUCLEON EOS (black lines), showing (from top left to bottom right) the total pressure, $P$, 
 selected mass fractions, $X_i$, corresponding chemical potentials, $\mu_i$ without restmass $m_i$ and the 
 effective masses, $m_i^*$, as well as the abundances of electrons and muons, $Y_e$ and $Y_\mu$, respectively. 
\label{fig:eos_a}}
\end{center}
\end{figure*}

\begin{figure*}[htp]
\begin{center}
\includegraphics[angle=0.,width=1\columnwidth]{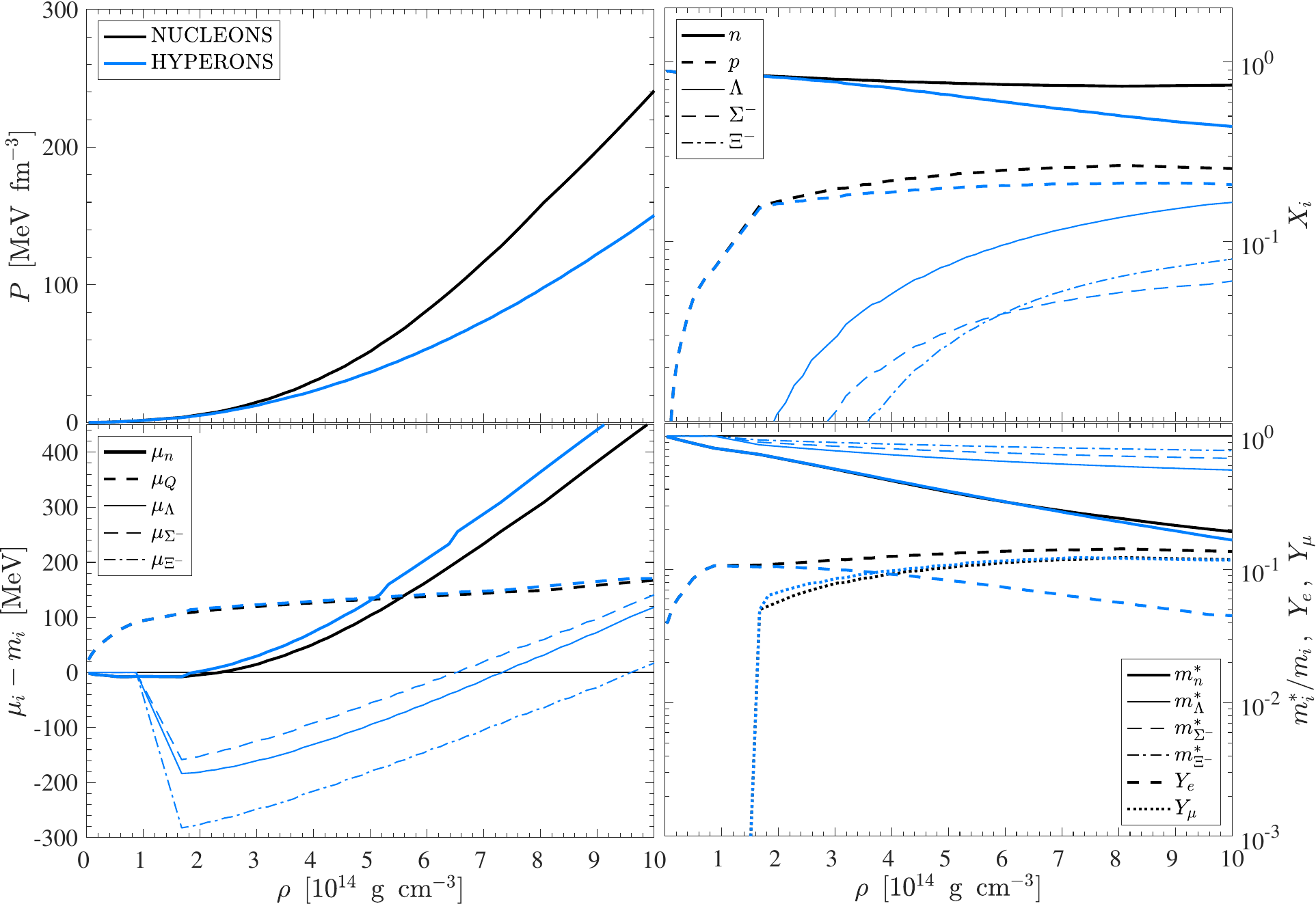}
\caption{~Same as Fig.~\ref{fig:eos_a} but for finite entropy per baryon of $s=3~k_{\rm B}$. 
\label{fig:eos_b}}
\end{center}
\end{figure*}

Figures~\ref{fig:eos_a} and \ref{fig:eos_b} illustrate selected quantities of the NUCLEONS EOS (black lines) and 
HYPERONS EOS (blue lines) at certain conditions, at $T=0$ in Fig.~\ref{fig:eos_a} and at a constant entropy of 
$s=3~k_{\rm B}$ in Fig.~\ref{fig:eos_b}, both assuming $\beta$-equilibrium. 
For both cases it becomes evident that the HYPERONS EOS is substantially softer at high densities than the NUCLEONS EOS,
illustrated by the pressure in the upper left panels of Figs.~\ref{fig:eos_a} and \ref{fig:eos_b}. 
This is a consequence from the lower neutron abundance since the EOS is dominated by the neutron degeneracy 
at the high densities explored here (see therefore the hadron abundances in the upper right panels and note the 
generally low abundances\footnote{
In the case of particles with baryon number equal one, abundances $Y_i$ and mass fractions $X_i$ are identical, defined through the partial densities $n_i=Y_i\,n_{\rm B}$ with baryon density $n_{\rm B}$.
} 
of hyperons $\Lambda$, $\Sigma^-$ and $\Xi^-$). 
Note also the modified charge neutrality conditions in the presence of hyperons, illustrated via the electron 
and  muon abundances, $Y_e$ and $Y_\mu$, respectively, in the lower right panels, together with the density 
dependence of the effective masses gap equations~\eqref{eq:meff}. 

The assumption of $\beta$-equilibrium results in the condition of equal chemical potentials for electrons and muons, 
$\mu_e=\mu_\mu$, once the electron chemical potential exceeds the muon restmass, $\mu_e\geq m_\mu=106$~MeV, 
which was employed here for the calculation of the muon abundances. 
Note that this assumption differs from the conditions realized in the CCSN simulations. 
This discrepancy has been pointed out in Refs.~\cite{Fischer2020PhRvD102,Fischer:2021}, 
that while $\mu_\mu\rightarrow\mu_e$, this condition can be reached only after more than 30~s post 
bounce in simulations of the PNS deleptonization and cooling, studied based on the DD2 relativistic mean field EOS, 
with density dependent meson-nucleon couplings \cite{Typel:2009sy}. 
As a consequence of this simplification applied here for the calculation of the muon EOS, the muon abundance 
$Y_\mu$ is found to be significantly larger than what we find in the CCSN and PNS simulations, that will be 
discussed in Sec.~\ref{sec:PNSdelept} below. 

The lower right panels in Figs.~\ref{fig:eos_a} and \ref{fig:eos_b} also show the effective hadron masses, 
$m_i^*$, relative to their vacuum values $m_i$, from which it becomes evident that the neutron mass decreases 
most rapidly towards about 15\% of its vacuum value  at a density of slightly above $\rho=10^{15}$~g~cm$^{-3}$, 
whereas the effective mass of the $\Lambda$ is about 50\% of its vacuum value at the same density.
The corresponding chemical potentials,  $\mu_i$ without rest masses $m_i$, are shown in the lower left panels of 
Figs.~\ref{fig:eos_a} and \ref{fig:eos_b}. 
At zero temperature, hyperons are strictly suppressed when their effective chemical potentials, 
shown in the lower left panel in Fig.~\ref{fig:eos_a}, are below their respective rest masses, while 
at finite temperature, shown in Fig.~\ref{fig:eos_b}, this condition does not apply any longer.

\section{Simulations of the PNS deleptonization with hyperons}
\label{sec:PNSdelept}
For the simulations of the PNS deleptonization, the {\tt AGILE-BOLTZTRAN} spherically symmetric general relativistic 
neutrino radiation hydrodynamics model is employed 
(c.f. Refs.~\cite{Mezzacappa93a,Mezzacappa93b,Mezzacappa93c,Liebendorfer04} and references therein). 
It is based on six-species Boltzmann neutrino transport and a complete set of weak interactions (the weak reactions 
used in the current study, including the references, are listed in Table~1 of ref.~\cite{Fischer2020PhRvC101}). 
We distinguish between $\mu$ and $\tau$ (anti)neutrinos due to the inclusion of weak reactions involving muons 
and antimuons. 
These include the muonic charged current reactions, based on the full kinematics approach \cite{Guo2020PhRvD102}, 
as well as neutrino-(anti)muon inelastic scattering. 
For both of which we are following the implementation of ref.~\cite{Fischer2020PhRvD102}. 

\begin{figure*}[htp]
\begin{center}
\subfigure[~18~M$_\odot$ progenitor]{\includegraphics[angle=0.,width=1.0\columnwidth]{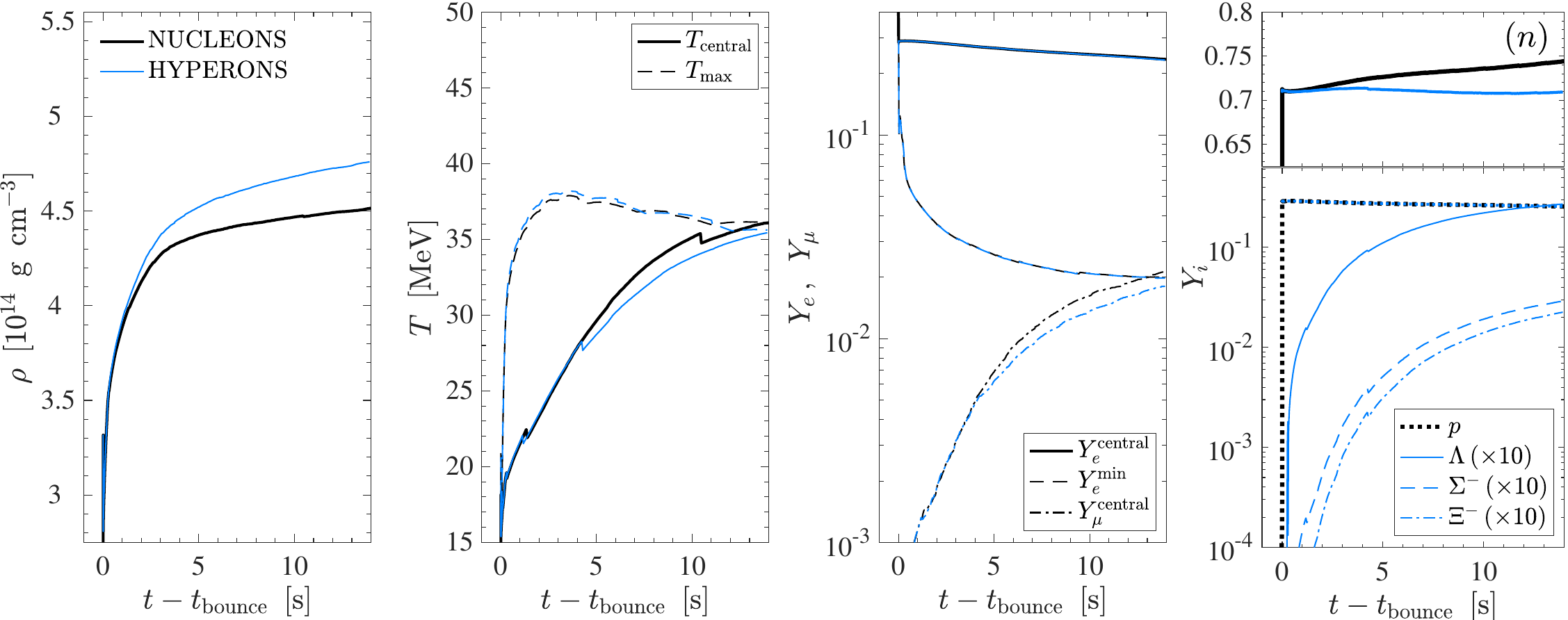}\label{fig:s18_central}}
\\
\subfigure[~25~M$_\odot$ progenitor]{\includegraphics[angle=0.,width=1.0\columnwidth]{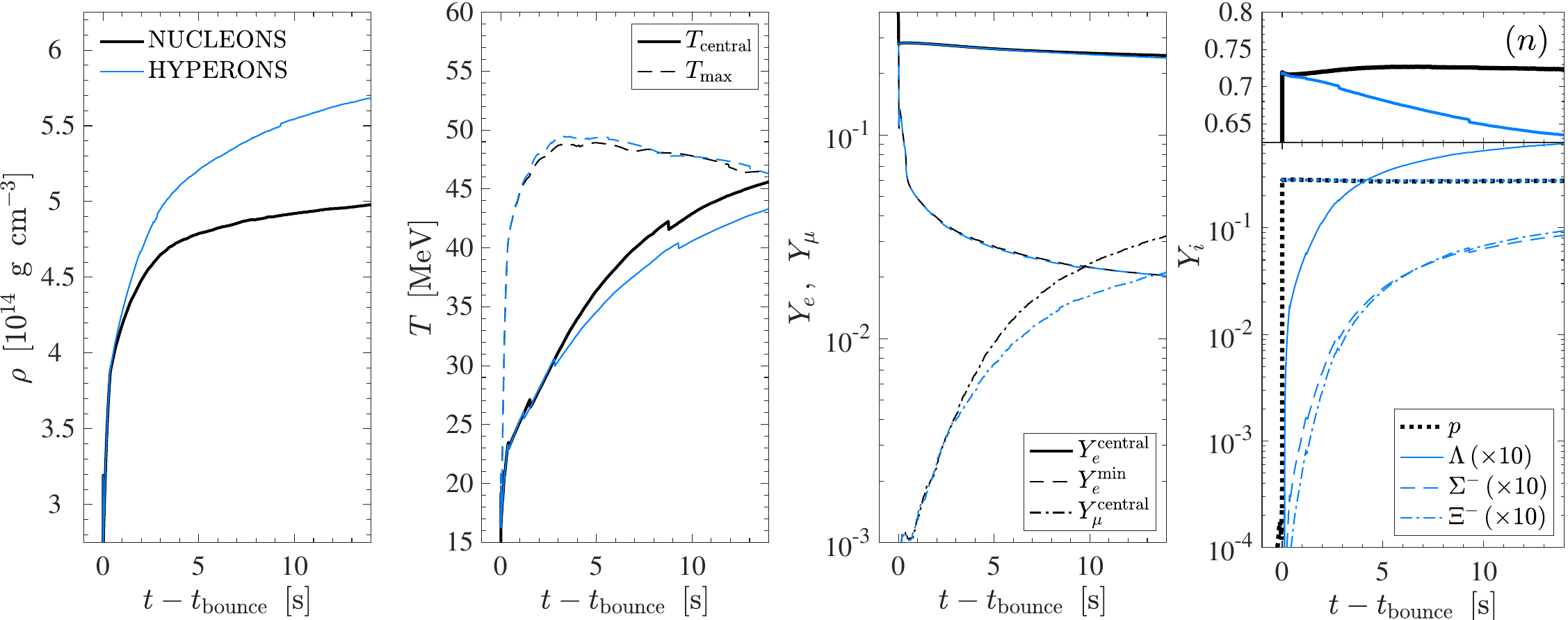}\label{fig:s25_central}}
\caption{~Post bounce evolution of selected central quantities for the 18~M$_\odot$ and 25~M$_\odot$ models in graph~(a) and (b) respevtively, comparing simulations with the NUCLEONS (black lines) and HYPERONS (blue lines) EOS showing from left to right, central restmass density $\rho$, central and maximum temperatures, $T_{\rm central}$ and $T_{\rm max}$, central electron and muon abundances, $Y_e$ and $Y_\mu$, and minimum $Y_e$, and the mass fractions $X_i$ of neutrons in the top panel  (thick solid lines) and in the bottom panel protons (thick dotted lines) in comparison to the strange hadrons $\Lambda$ (thin solid lines), $\Sigma^-$ (thin dashed lines) and $\Xi$ (thin dash-dotted lines). Note that the latter are magnified by one order of magnitude for better visibility.}
\end{center}
\label{fig:evol_central}
\end{figure*}

{\tt AGILE-BOLTZTRAN} has a flexible EOS module that contains the Lattimer \& Swesty Skyrme model \cite{LSEOS} 
as well as the comprehensive RMF EOS catalogue of ref.~\cite{Hempel12}. 
The latter incorporates the modified nuclear statistic equilibrium (NSE) with several thousand nuclei \cite{Hempel2010NuPhA837}, 
including excited state contributions, modelled via temperature dependent statistical weight factors for each nuclear species. 
The transition from the modified NSE to homogeneous nuclear matter is taken into account via a first-order phase transition 
construction where the nuclear states are suppressed through a geometric excluded volume approach 
(for alternative excluded volume functionals, c.f. ref.~\cite{Fischer2016EPJA52} and references therein). 
In ref.~\cite{Hempel12}, the excluded volume parameter is chosen to result in the complete disappearance of all nuclei 
at nuclear saturation density. 
In the present study, we replace the supersaturation density phase of the DD2 RMF EOS \cite{Typel:2009sy,Hempel12}, 
with density-dependent meson-nucleon mean-field couplings, with the NUCLEONS and HYPERONS settings from FSU2H$^*$. 
We find that this approach, despite its ad hoc nature, results in a smooth transition. 
EOS contributions from electrons, positron and photons, as well as Coulomb contributions for the non-NSE regime,
are implemented based on the routines of ref.~\cite{Timmes1999}.

The inclusion of the HYPERONS EOS is connected with the implementation of the associated hyperon degrees of freedom, 
$\Lambda$, $\Sigma^-$ and $\Xi^-$, into the {\tt AGILE-BOLTZTRAN} CCSN model. 
It concerns their abundances as well as the single particle properties such as the effective masses, the chemical potentials 
and the mean-field potentials, according to the underlying model as was discussed in Sec.~\ref{sec:eos}. 
In particular, the latter will become relevant when implementing weak processes involving hyperons, which, 
for the present study, are being omitted as the simulation times are limited to $\mathcal{O}(10~{\rm s})$ 
when neutrinos still decouple at lower densities where no hyperons are abundant. 
This situation will change towards later times, when neutrinos start to decouple in the regions where hyperons are abundant, 
such that the inclusion of weak processes involving hyperons would influence the further deleptonization and cooling behaviour. 
Furthermore, the presence of hyperons modifies the charge-neutrality condition, with the negatively charged hyperons 
considered, i.e. $Y_e+Y_\mu = Y_p - Y_{\Sigma^-} - Y_{\Xi^-}$. 
Note that 
despite their consistent treatment and implementation into the CCSN simulations, 
the heavier, neutral hyperons, 
$\Sigma^0$ and $\Xi^0$, and the positively charged hyperons, $\Sigma^+$ and $\Xi^+$, are neglected in the following discussions, 
as they are all strongly Boltzmann suppressed under the extremely neutron-rich conditions 
at highest densities encountered at the PNS interior, i.e. the following abundance hierarchy generally holds:
$Y_\Lambda > Y_{\Sigma^-} > Y_{\Xi^-}$ 
as well as 
$Y_{\Sigma^-} \gg Y_{\Sigma^0} > Y_{\Sigma^+}$ 
and 
$Y_{\Xi^-} \gg Y_{\Xi^0} > Y_{\Xi^+}$. 

With this enhanced EOS module setup, simulations of CCSN are launched from two progenitor models, 
with zero-age main sequence masses of 18~M$_\odot$ and 25~M$_\odot$, both from the stellar evolution 
series of ref.~\cite{Woosley:2002zz}. 
These stellar progenitors are being evolved through all CCSN phases self consistently. 
It is found that up to several 100~ms post bounce, the abundance of hyperons remains small, 
below 1\%, and no impact could be observed, 
not on the overall dynamics nor on the neutrino heating and cooling rates. 
The neutrino signals are indistinguishable between NUCLEONS and HYPERONS.
 
The simulations with the NUCLEONS and HYPERONS EOS proceed identically. 

Since neutrino-driven explosions cannot be obtained for such massive iron-core progenitors
in spherically symmetric CCSN simulations, we follow closely the procedure outlined in 
ref.~\cite{Fischer10}. 
Therefore, the electronic charged current weak rates are being enhanced in the gain layer, 
which results in the continuous shock expansion and the onset of the CCSN explosion. 
Once the shock reaches a radius of about 1000~km, we switch back to the standard weak 
rates, following the later PNS deleptonization. 
A similar procedure has been employed in ref.~\cite{Jost2024arXiv240714319J} for a
large sample of CCSN explosion models for various different progenitors at different 
metalicities.

\begin{table}
\begin{center}
\caption{Stellar progenitor model's zero-age main sequence masses, $M_{\rm ZAMS}$, and the encloses PNS baryonic and gravitational masses, $M_{\rm B}$, and $M_{\rm G}$, as well as the corresponding radii $R$, evaluated at the end of the CCSN simulations.}
\begin{tabular}{|lllllllllllll|}
\hline
progenitor mass$^{\rm a}$ & && EOS & && $M_{\rm B}^{\rm b}$& && $M_{\rm G}^{\rm b}$ & && $R^{\rm b}$ \\
$M_{\rm ZAMS}\,[$M$_{\odot}]$ & && && & $[$M$_{\odot}]$ & && $[$M$_{\odot}]$ & && $[$km$]$ \\
\hline
18 & && NUCLEONS & && 1.543 & && 1.435 & && 14.18 \\
18 & && HYPERONS & && 1.558 & && 1.450 & && 14.09 \\
25 & && NUCLEONS & && 1.942 & && 1.792 & && 14.84 \\
25 & && HYPERONS & && 1.938 & && 1.791 & && 14.45\\
\hline
\end{tabular}
\end{center}
Notes.\\
$^{\rm a}$~models from the stellar evolution series of ref.~\cite{Woosley:2002zz} \\
$^{\rm b}$~Evaluated at $\rho=10^{10}$~g~cm$^{-3}$ at about 10~s post bounce
\label{table:runs}
\end{table}

\begin{figure}[t!]
\begin{center}
\includegraphics[angle=0.,width=0.65\columnwidth]{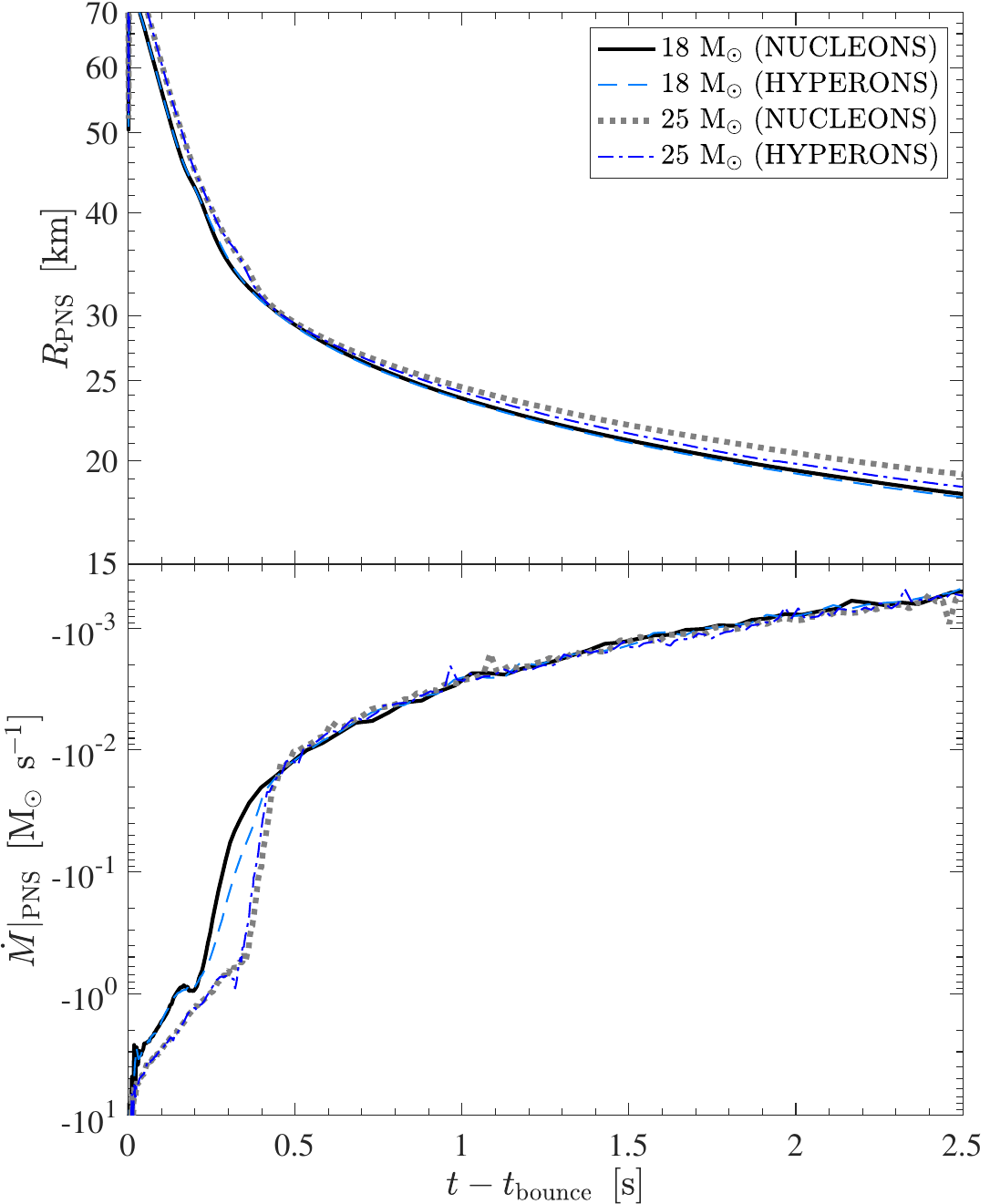}
\caption{~Post bounce evolution of the PNS radius $R_{\rm PNS}$ (top panel) and mass accretion rate $\dot M\vert_{\rm PNS}$ (bottom panel), both evaluated at a selected restmass density of $\rho=10^{11}$~g~cm$^{-3}$, comparing the 18~M$_\odot$ simulations for NUCLEONS (black solid lines) and HYPERONS (light blue dashed lines) and the 25~M$_\odot$ simulations also for NUCLEONS (dark grey solid lines) and HYPERONS (dark blue dashed lines). \label{fig:R_mdot}}
\end{center}
\end{figure}

The post-bounce evolution of the four models is illustrated in Fig.~\ref{fig:evol_central}, 
for 18~M$_\odot$ and 25~M$_\odot$ model in Figs.~\ref{fig:s18_central} and \ref{fig:s25_central}, 
respectively, distinguishing the NUCLEONS (black lines) and HYPERONS (blue lines) EOS. 
Once the supernova explosion proceeds with the revival of the shock, mass accretion onto 
the PNS surface ceases and the enclosed PNS baryon masses change only marginally. 
The latter is due to the ejection of the neutrino driven wind, initially on the order of 
$10^{-3}$~M$_\odot$~s$^{-1}$ and later substantially lower with $10^{-4}$~M$_\odot$~s$^{-1}$. 
The resulting PNS masses of the two sets of simulations are listed in Table~\ref{table:runs}, 
for both NUCLEONS and HYPERONS runs. 
These values are evaluated at the restmass density of $10^{11}$~g~cm$^{-3}$ at about 
10~s post bounce. 
The larger PNS masses for the more massive progenitor are due to the higher post-bounce 
mass accretion rates for the 25~M$_\odot$ models, in particular during the early post-bounce 
phase, i.e., before the shock revival, as is illustrated in the bottom panel of Fig.~\ref{fig:R_mdot},
showing the mass accretion rate $\dot M_{\rm PNS}$, as well evaluated at $\rho=10^{11}$~g~cm$^{-3}$. 
The top panel illustrates the different PNS radii, $R_{\rm PNS}$, and their evolution 
for all simulations with the NUCLEONS and HYPERONS EOS under investigation, 
showing the slightly faster contraction of the HYPERONS PNS due to the softer 
supersaturation density behaviour. 

Only after several seconds, in both simulations, differences due to the inclusion of hyperons arise. 
These differences are best seen by the evolution of the central densities, shown in the left panels of 
Figs.~\ref{fig:s18_central} and \ref{fig:s25_central}, where the simulations based on the HYPERONS EOS reach significantly higher values than the ones with NUCLEONS. 
This EOS softening is due to  the continuous rise of the hyperon abundances for $\Lambda$, $\Sigma^-$ 
and $\Xi^-$, shown in the right panels. 
Note that these hyperons are absent for the NUCLEONS simulations (black lines). 
The first hyperons which appear are $\Lambda$'s, due to the lowest mass (see Figs.~\ref{fig:eos_a} 
and \ref{fig:eos_b}), already at a density slightly in excess of nuclear saturation density, i.e. already 
during the early post-bounce evolution prior to the supernova explosion onset. 
However, the abundance of $\Lambda$ remains low, on the order of $Y_\Lambda\simeq 0.001-0.025$ 
for the for 18~M$_\odot$ model and $Y_\Lambda\simeq 0.01-0.07$ for the for 25~M$_\odot$ model. 
The latter reach higher abundances because of the generally higher densities and temperatures 
obtained at the central fluid elements in the simulations. 
Note therefore the different evolution of the central and maximum temperatures in Figs.~\ref{fig:s18_central} 
and \ref{fig:s25_central}. 
The abundances of the other hyperons, $\Sigma^-$ and $\Xi^-$, remain about one order of magnitude lower than those of $\Lambda$, in both simulations. 

\begin{figure*}[htp]
\begin{center}
\subfigure[~18~M$_\odot$ luminosities]{\includegraphics[angle=0.,width=0.48\columnwidth]{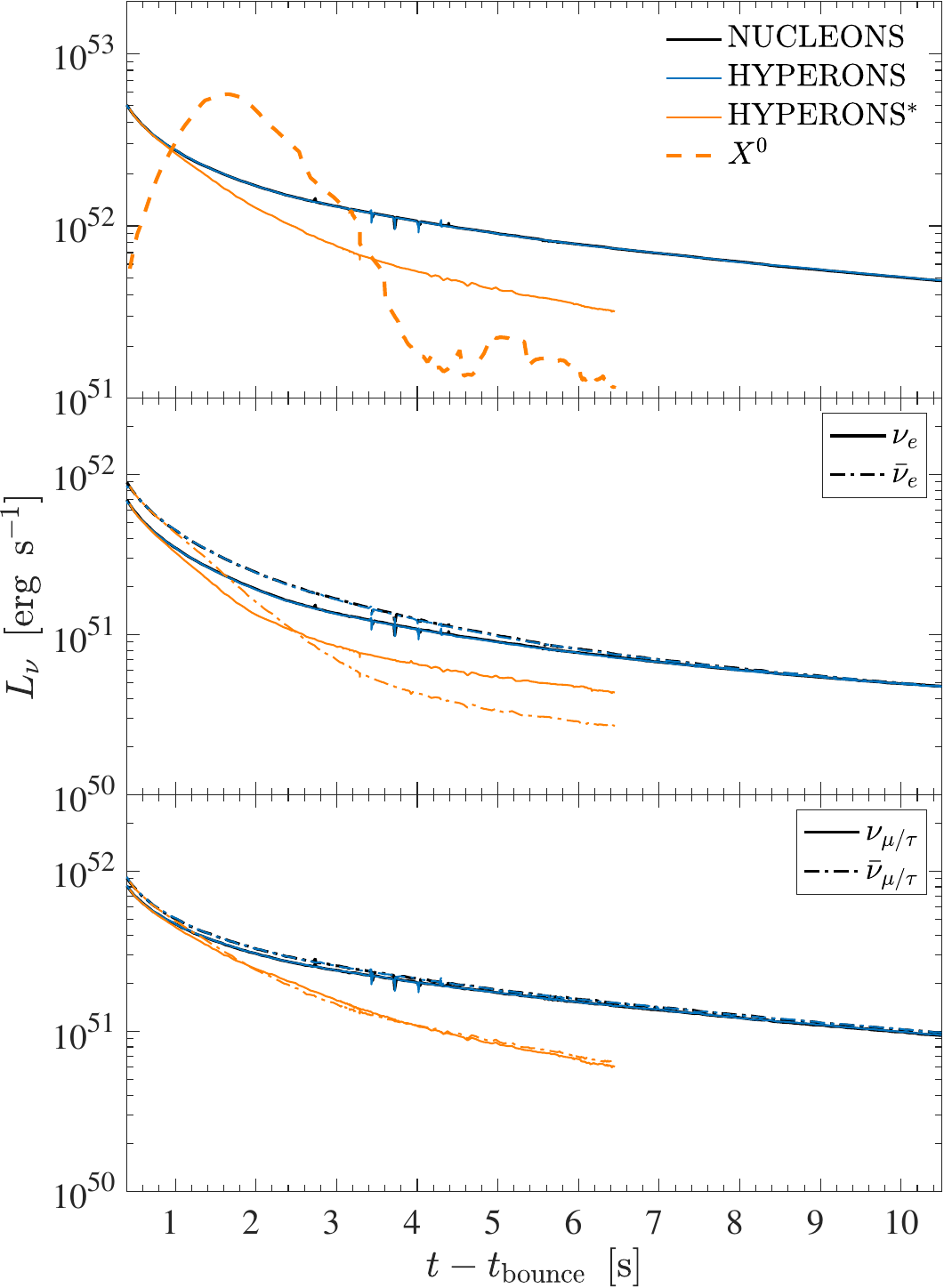}
\label{fig:s18_lumin}}
\hfill
\subfigure[~25~M$_\odot$ luminosities]{\includegraphics[angle=0.,width=0.48\columnwidth]{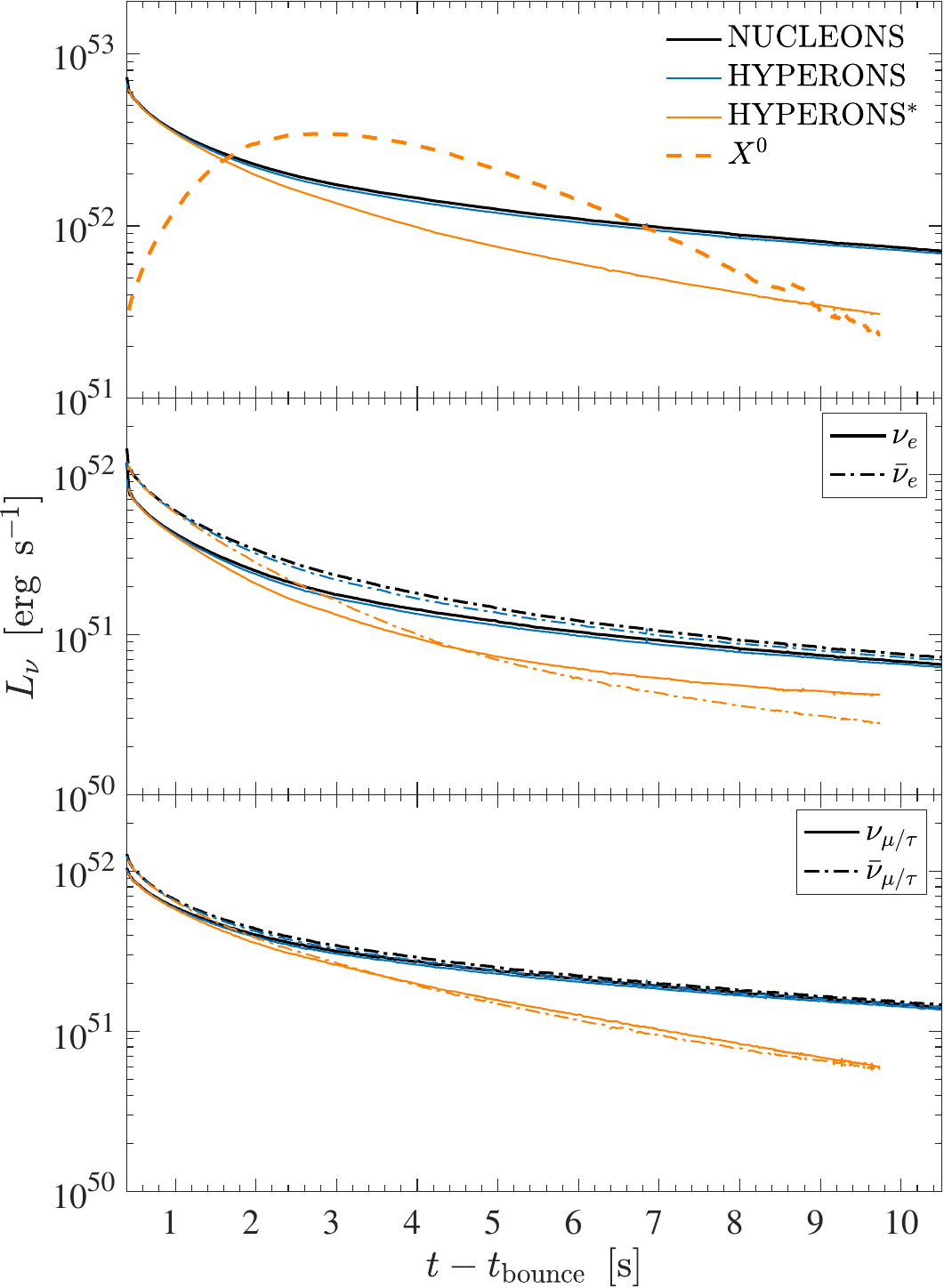}\label{fig:s25_lumin}}
\\
\subfigure[~18~M$_\odot$ average neutrino energies]{\includegraphics[angle=0.,width=0.48\columnwidth]{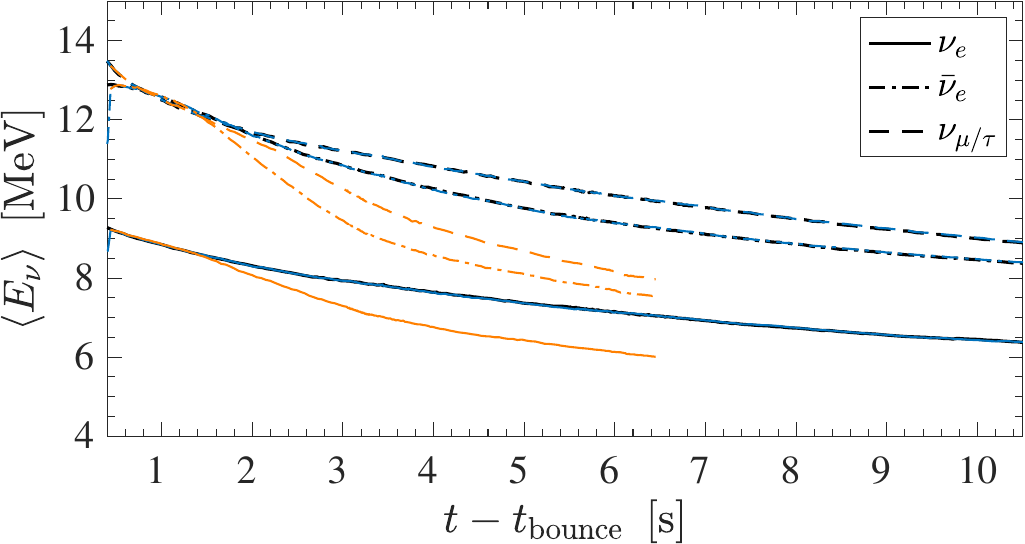}
\label{fig:s18_rms}}
\hfill
\subfigure[~25~M$_\odot$ average neutrino energies]{\includegraphics[angle=0.,width=0.48\columnwidth]{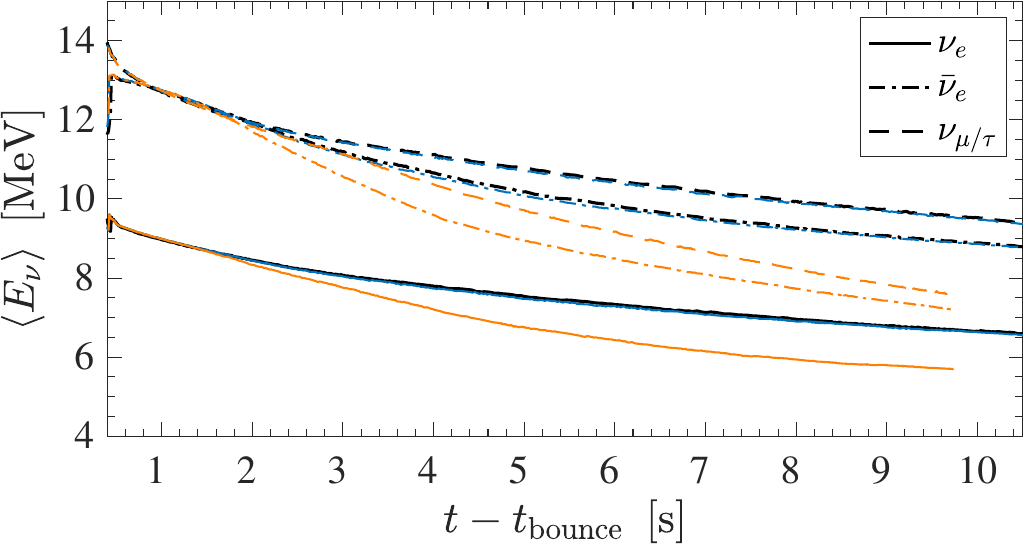}\label{fig:s25_rms}}
\caption{~Post bounce evolution of the neutrino luminosities and average energies for the two sets of simulations launched from 18~M$_\odot$ (left panel) and 25~M$_\odot$ progenitors (right panel), comparing in addition to the NUCLEONS (black lines) and HYPERONS (blue lines) EOS also those with the inclusion of losses through the emission of dark sector bosons $X_0$, denoted as HYPERONS$^*$ (orange lines). The luminosity in graphs~(a) and (b) distinguish the total neutrino luminosity (top panels, solid lines) in comparison to the dark sector particle, denoted as $X_0$ (dashed lines), electron flavours (middle panels) and heavy lepton flavours (bottom panel). The average neutrino energies in graphs~(c) and (d) show the $\nu_e$ (solid lines), $\bar\nu_e$ (dash dotted lines) and $\nu_{\mu/\tau}$ (dashed lines) as representative for all heavy lepton flavours. \label{fig:lumin}}
\end{center}
\end{figure*}

Another difference between the runs with NUCLEONS and HYPERONS is the feedback to the nucleons. 
As was already discussed in Sec.~\ref{sec:eos}, in particular the abundance of neutrons is reduced  due 
to the appearance of $\Lambda$, effectively replacing neutrons with $\Lambda$'s. 
This phenomenon is observed consistently in the CCSN simulations too. 
The neutron abundances are shown in the right panels of Figs.~\ref{fig:s18_central} and \ref{fig:s25_central}. 
The reduced neutron abundance in addition softens the high-density EOS, which is governed  
by neutrons with a dominating abundance of $Y_n\simeq 0.75$--$0.7$. 
As a consequence of the different hadronic abundances, in order to obey the condition 
of charge neutrality, there is a feedback to the electron and muon abundances. 
In particular, the central muon abundance is reduced substantially for the massive 
25~M$_\odot$ model. 
Note also that the central muon abundance is substantially lower and that the central 
electron abundance is substantially higher than in the EOS discussion in Sec.~\ref{sec:eos}, 
where $\beta$-equilibrium was assumed. 
The observed difference of the electron and in particular the muon abundances are a 
consequence of the lower central temperatures obtained for the simulations based on 
the HYPERONS EOS (see Fig.~\ref{fig:evol_central}), since the muons are produced 
thermally through the muonic charged current processes. 

Despite the differences obtained at the PNS interior, structure and stability of the PNS 
are not affected by the presence of hyperons. 
This is partly related to the conditions for the appearance of hyperons, which in turn 
gives rise to finite hyperon abundances, in particular $\Lambda$ only at the very 
central fluid elements in the simulations. 
As a consequence, the locations of the neutrinospheres are not affected by the 
presence of hyperons and hence the neutrino luminosities and average energies 
are indistinguishable between the runs NUCLEONS and HYPERONS,
for the intermediate-mass 18~M$_\odot$ model, shown in Figs.~\ref{fig:s18_lumin} 
and \ref{fig:s18_rms}, and high-mass 25~M$_\odot$ model, 
shown in Figs.~\ref{fig:s25_lumin} and \ref{fig:s25_rms}. 
The evolution of the neutrino luminosities and average energies shown in 
Fig.~\ref{fig:lumin} are sampled in the co-moving frame of reference at a distance of 
500~km.

\section{Dark flavoured sectors during PNS deleptonization}
\label{sec:DM}
The results including the dark flavoured sector are based on the same HYPERON EOS 
and will be henceforth denoted as HYPERONS$^*$, which will be discussed in this section. 
To investigate the impact of a dark flavoured sector in CCSN, we need to estimate the 
emissivity in the new channels. 
We assume that the emitted particles are massless (i.e. their mass is much smaller 
than $T_{\rm PNS}$) and that the interactions are such that their mean free path is 
much larger than the radius of the PNS. 
These particles will then freely stream out of the PNS once they are produced in its core.

\subsection{Dark flavoured emissivity}
For concreteness, in the following we will focus on the emission of neutral dark bosons $X_0$, 
such as an axion or dark photon, produced by the flavour-violating transition $s\to d + X_0$. 
This is parametrised by the following Lagrangians,
\begin{eqnarray}
\label{eq:Lag_axion}
\mathcal L_a 
&=& 
\frac{1}{2f_a}
\left(\partial_\mu a\right)
{\bar {\rm d}}\, \gamma^\mu(c^{V}_{\rm ds}
+
c^{A}_{\rm ds}\gamma_5)\, {\rm s}+{\rm h.c.}~,
\\
\label{eq:Lag_DP}
\mathcal L_{\gamma^\prime} 
&=&
\frac{1}{\Lambda} {\bar {\rm d}}\,\sigma^{\mu\nu}(c^{T}_{\rm ds}
+
c^{T5}_{\rm ds}\gamma_5)
\,
{\rm s}\,F_{\mu\nu}^\prime+{\rm h.c.}~,
\end{eqnarray}
where $a$ is the axion field and $f_a$ its decay constant and with strange and down quark 
fields denoted as ${\rm s}$ and ${\bar{\rm d}}$, respectively. 
$F_{\mu\nu}^\prime$ is the field strength tensor of the dark photon and $\Lambda$ the 
energy scale associated to the UV completion of the effective Lagrangian 
$\mathcal L_{\gamma^\prime}$ 
(see ref.~\cite{Eguren:2024oov} for effective vectorial couplings of the dark photon and some UV completions), and $c^{V,A}_{\rm ds}$ and $c^{T,T5}_{\rm ds}$ are 
dimensionless (effective) couplings. 
The main hadronic production mechanism in the PNS induced by Lagrangians
\eqref{eq:Lag_axion} and \eqref{eq:Lag_DP} will then be $\Lambda\to n + X_0$, 
whereby decays involving heavier hyperons, such as $\Sigma^+\to p + X_0$, 
are suppressed by their lower abundances and neglected in the present analysis. 
Following ref.~\cite{MartinCamalich:2020dfe}, the spectrum of the energy loss rate 
of the nuclear medium per unit volume, denoted as $Q$, with respect to the $X_0$ 
energy $E_{a}$, in the PNS rest frame and induced by the process $\Lambda\to n + X_0$ 
is given by the following expression,
\begin{equation}
\label{eq:differentialQ}
\frac{dQ}{dE_a}
=
\frac{m_\Lambda^2\Gamma_X E_a}{2\pi^2{\bar E}}\int_{E_0}^\infty dE \, f_\Lambda(1-f_n)~, 
\end{equation}
with ${\bar E}=(m_\Lambda^2-m_n^2)/2m_\Lambda$, $E$ is the energy of the 
$\Lambda$ with $E_0=m_\Lambda(E_a^2+{\bar E}^2)/(2 E_a {\bar E})$ and
$f_{\Lambda,n}$ are the relativistic Fermi-Dirac phase-space distribution functions 
of $\Lambda$ and $n$. 
The information of the transition amplitude is encoded in $\Gamma_X$, which is 
the decay rate of $\Lambda\to n + X_0$ in vacuum,
\begin{equation}
\label{eq:rate}
\Gamma_X=\frac{\bar E^3 C_X}{2\pi}~,
\end{equation}
with $C_a=(f_1^2|c_{\rm ds}^V|^2+g_1^2|c_{\rm ds}^A|^2)/4f_a^2$, where $f_1=-1.22(6)$ 
and $g_1=-0.89(2)$ are the vector and axial-vector baryonic couplings determined in 
ref.~\cite{MartinCamalich:2020dfe}, and 
$C_{\gamma^\prime}=8 g_T^2(|c_{\rm ds}^T|^2+|c_{\rm ds}^{T5}|^2)/\Lambda^2$, 
where 
$g_T\approx-0.73$ is the baryonic tensor coupling 
(further details can be found in ref.~\cite{Camalich:2020wac} and references therein).  

The total emission rate can be estimated analytically by taking the nonrelativistic limit 
of the baryons, expanding to leading order in $\delta=(m_\Lambda-m_n)/m_n$ and 
neglecting the neutron's Pauli blocking \cite{MartinCamalich:2020dfe}. 
Expressed in terms of the emissivity, $\epsilon=Q/\rho$, one obtains the following result,
\begin{equation}
\epsilon \approx  \delta\;Y_\Lambda \tau_\Lambda^{-1}\, {\rm BR}(\Lambda\to n + X_0)~, 
\label{eq:emissivity_proxy}
\end{equation}
where we have re-expressed
the decay width in terms of the lifetime, $\tau_\Lambda$, and the branching fraction of the 
decay, {\rm BR}. 
Adopting the classical upper limit of the emissivity, 
$\epsilon_{\rm max}=10^{19}$~erg~s$^{-1}$ g$^{-1}$, 
from SN1987A at the conditions of the PNS predicted at about 1 second post-bounce, 
one obtains, 
\begin{align}
\epsilon\approx\epsilon_{\rm max}\left(\frac{Y_\Lambda}{0.01}\right)\left(\frac{1.6\times10^{-9}}{\text{BR}(\Lambda\to n + X_0)}\right)~.
\end{align}
This leads, for a characteristic $\Lambda$ abundance of 1\%, to an upper limit of the branching 
fraction in the range of $10^{-9}$, which overestimates the conservative limit by a 
factor $\approx 5$, which was obtained using the full expression in Eq.~\eqref{eq:differentialQ} 
and state-of-the-art spherically symmetric simulations reported in ref.~\cite{Bollig:2020xdr}. 
Nevertheless, this upper bound is many orders of magnitude stronger than those that can be 
obtained from laboratory experiments, and, as was found in 
Refs.~\cite{MartinCamalich:2020dfe,Camalich:2020wac}, it leads to the strongest constraint 
on the axial couplings of the flavoured QCD axion and on the massless dark photon couplings. 
In addition, as discussed in ref.~\cite{Camalich:2020wac}, in the case of $\Lambda$ processes 
one would still produce a too large emission in the deep trapping regime as the dark sector 
luminosity would stem from the last surface where $\Lambda$'s can coexist in equilibrium in the 
plasma and which corresponds to a region of high temperatures. 

\subsection{Impact of dark flavoured sector on PNS evolution}
\begin{figure*}[t!]
\begin{center}
\subfigure[~18~M$_\odot$: 1~s post bounce]{\includegraphics[angle=0.,width=0.495\columnwidth]{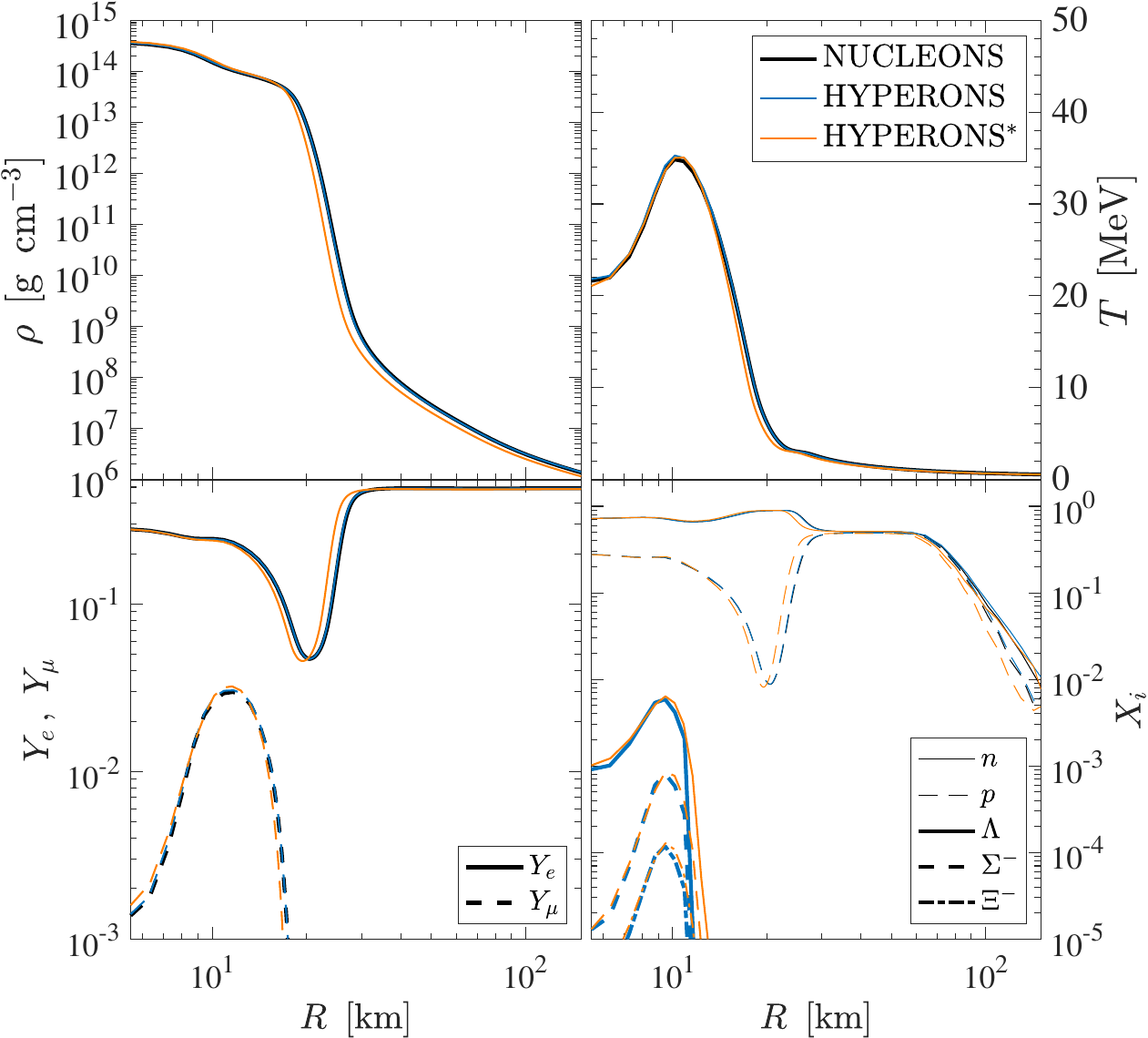}\label{fig:s18_hydro_a}}
\hfill
\subfigure[~25~M$_\odot$: 1~s post bounce]{\includegraphics[angle=0.,width=0.495\columnwidth]{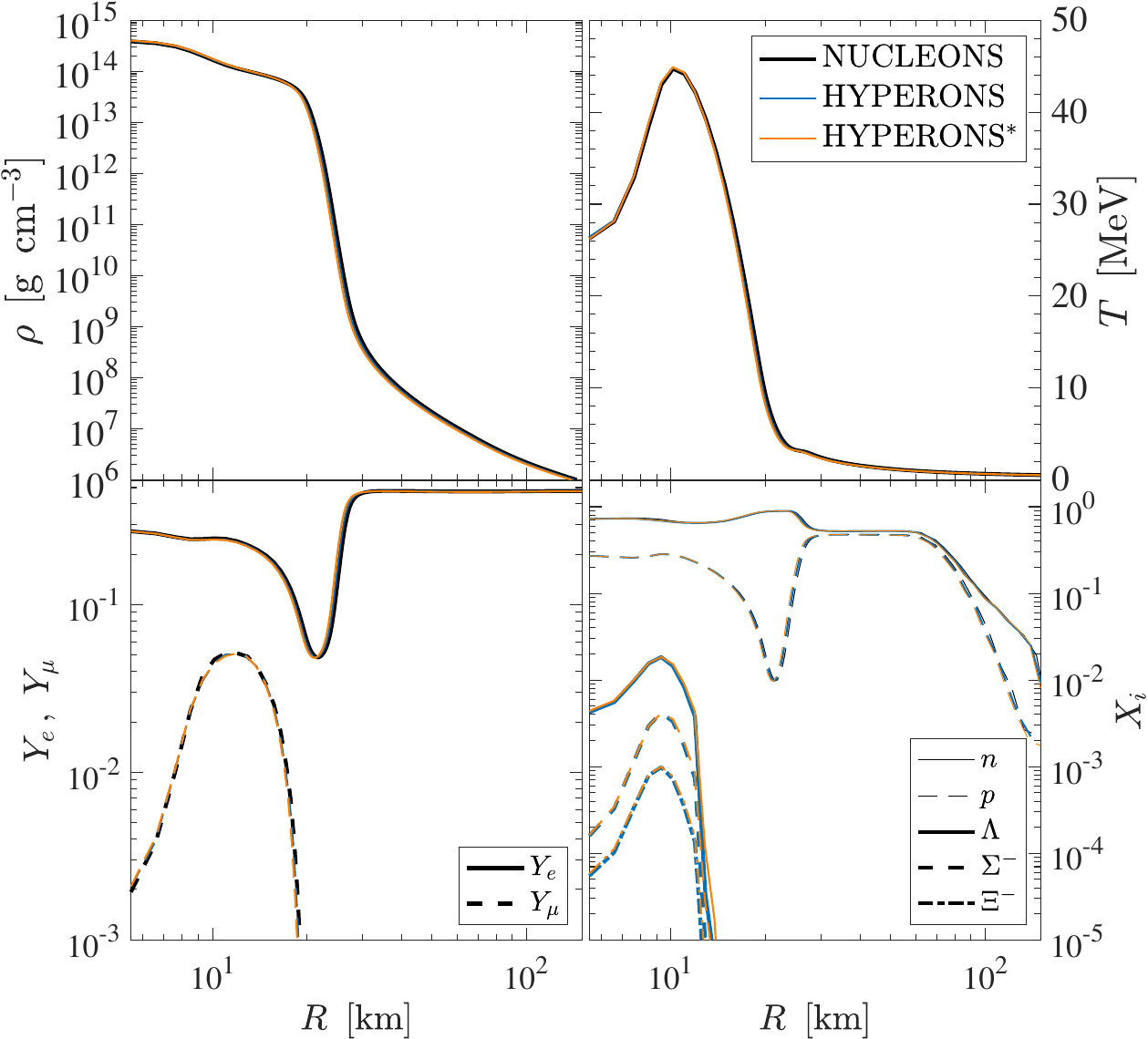}\label{fig:s25_hydro_a}}
\\
\subfigure[~18~M$_\odot$: 5~s post bounce]{\includegraphics[angle=0.,width=0.495\columnwidth]{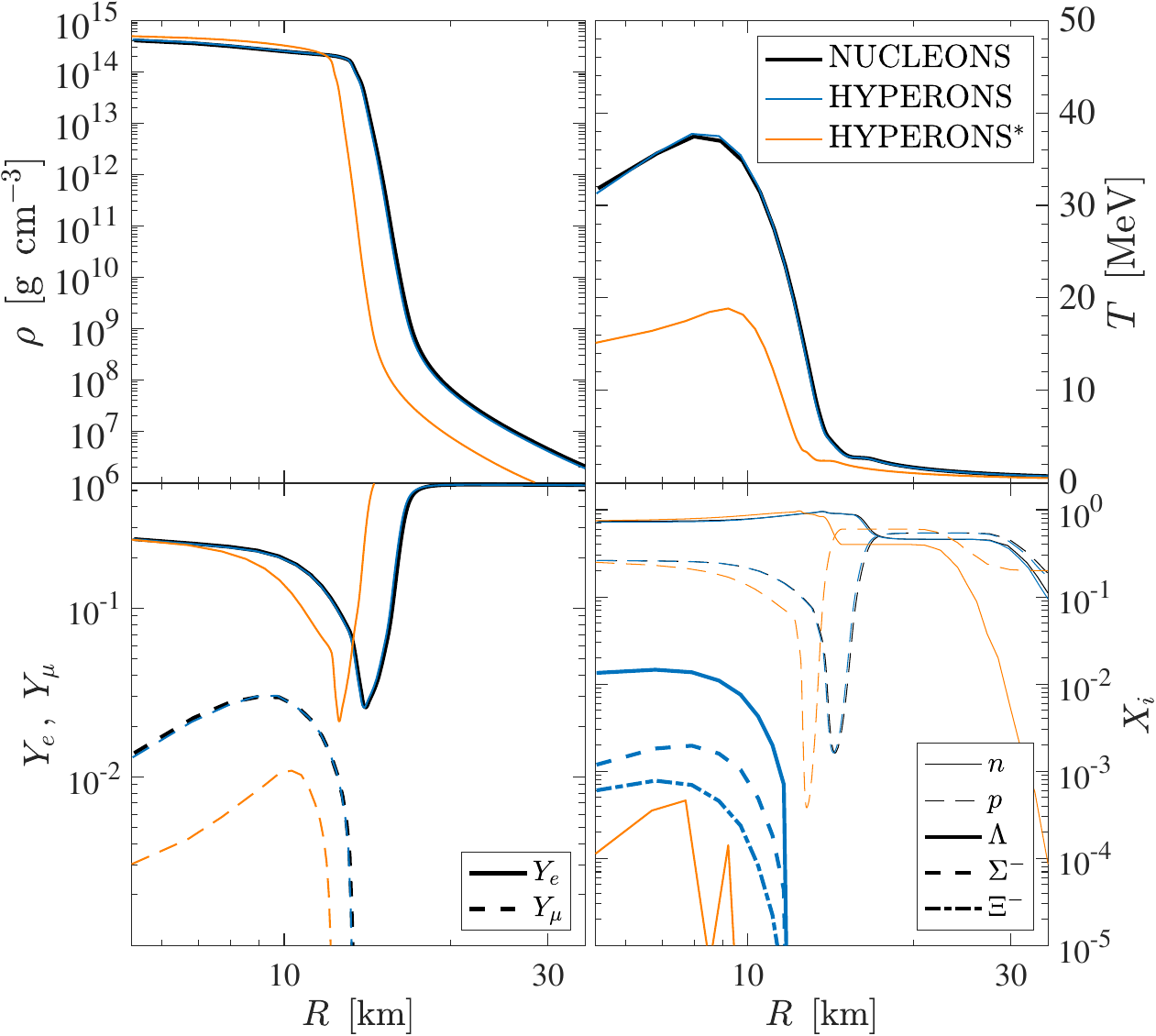}\label{fig:s18_hydro_b}}
\hfill
\subfigure[~25~M$_\odot$: 5~s post bounce]{\includegraphics[angle=0.,width=0.495\columnwidth]{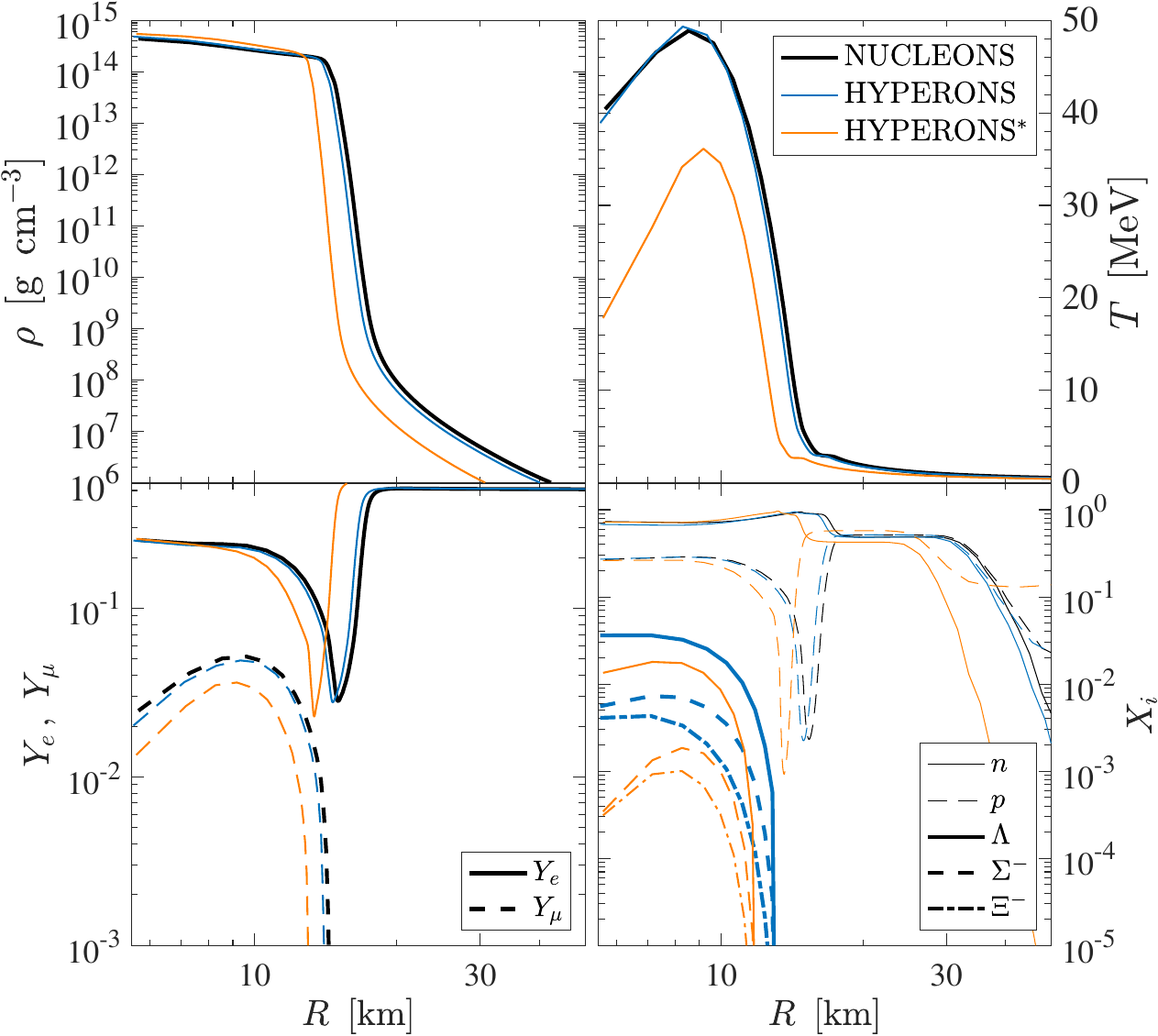}\label{fig:s25_hydro_b}}
\caption{~Radial profiles of selected quantities, showing the restmass density $\rho$ (top left panels), 
temperature $T$ (top right panels), electron and muons abundances, $Y_e$ and $Y_\mu$ (lower left panels), 
as well as the  mass fractions for neutrons ($n$), protons ($p$) and the hyperons $\Lambda$, $\Sigma^-$ and 
$\Xi^-$ (lower right panels), at two different post bounce times during the PNS deleptnoization, for the two sets 
of simulations launched from 18~M$_\odot$ (left panels) and 25~M$_\odot$ progenitors (right panels), 
comparing in addition to the NUCLEONS (black lines) and HYPERONS (blue lines) EOS also those with 
the inclusion of $X_0$ losses, denoted as HYPERONS$^*$ (orange lines). \label{fig:hydro}}
\end{center}
\end{figure*}

The non-relativistic emissivity~\eqref{eq:emissivity_proxy} is implemented into the 
{\tt AGILE-BOLTZTRAN} CCSN model as sink for the internal energy equation, following the 
implementation of dark boson losses of Refs.~\cite{Fischer2016PhRvD94_axions,Fischer:2021jfm}. 
Here, we use the literature branching ratios of $10^{-8}$ for the intermediate-mass 
18~M$_\odot$ model and $10^{-9}$ for the massive 25~M$_\odot$ model, in accordance 
with ref.~\cite{Camalich:2020wac}.
It results in enhanced cooling contributions in the domain where $\Lambda$ are 
abundant~\footnote{
These limits can be readily transformed in stringent limits on the flavoured QCD axion or 
massless dark photon described by the Lagrangians in~\eqref{eq:Lag_axion} and 
\eqref{eq:Lag_DP}. 
For example, in the QCD axion case with $\mathcal O(1)$ off-diagonal flavour couplings, 
the upper limit $\text{BR}(\Lambda\to n + a)\leq10^8$ corresponds to the limit 
$f_a\gtrsim5\times10^9\text{ GeV}$ or $m_a\lesssim 1\text{ meV}$ 
(see Refs.~\cite{Camalich:2020wac} and \cite{MartinCamalich:2020dfe} for details).
}.
This is the case only at the innermost 15--20~km of the PNS, as is illustrated for both 
18~M$_\odot$ and 25~M$_\odot$ models in Fig.~\ref{fig:hydro} for two distinct post-bounce 
times of 1~s and 5~s, comparing the three setups NUCLEONS (black lines), 
HYPERONS (blue lines) and  HYPERONS$^*$ (orange lines). 

\begin{figure}[t!]
\begin{center}
\subfigure[~18~M$_\odot$]{\includegraphics[angle=0.,width=0.495\columnwidth]{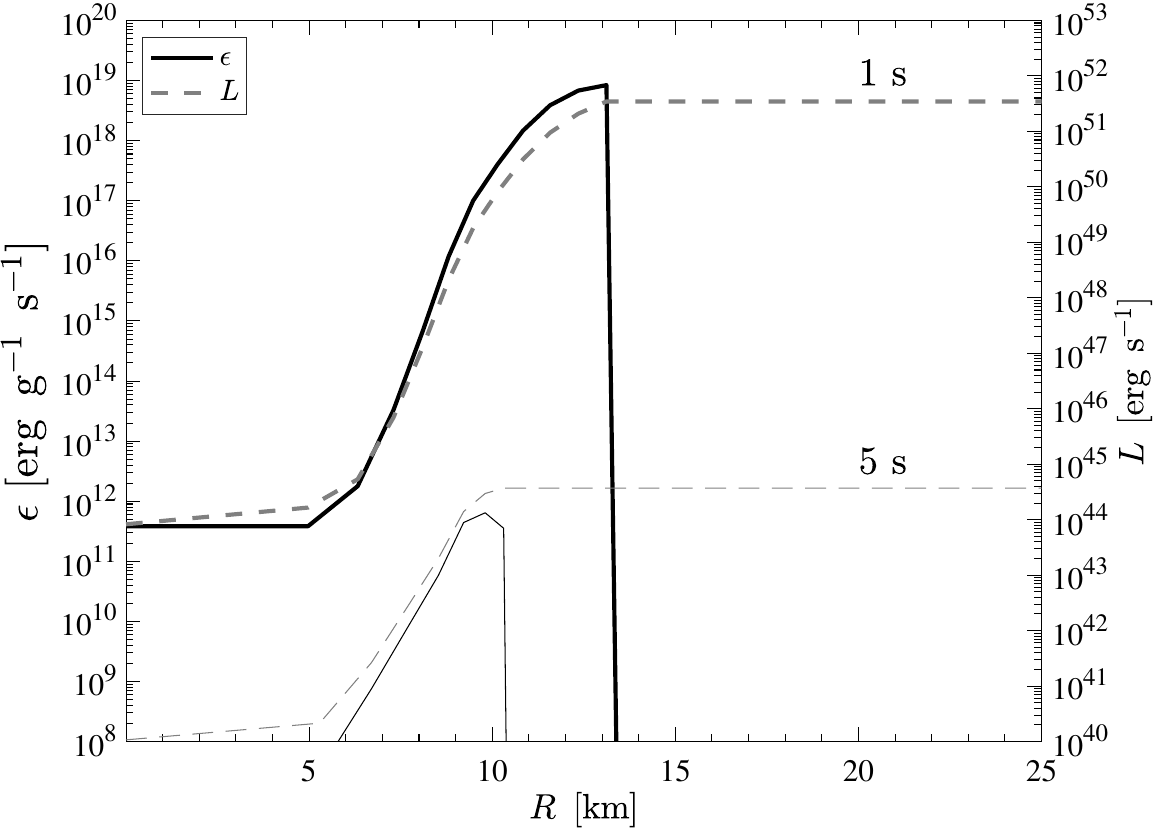}\label{fig:s18_dm}}
\subfigure[~25~M$_\odot$]{\includegraphics[angle=0.,width=0.495\columnwidth]{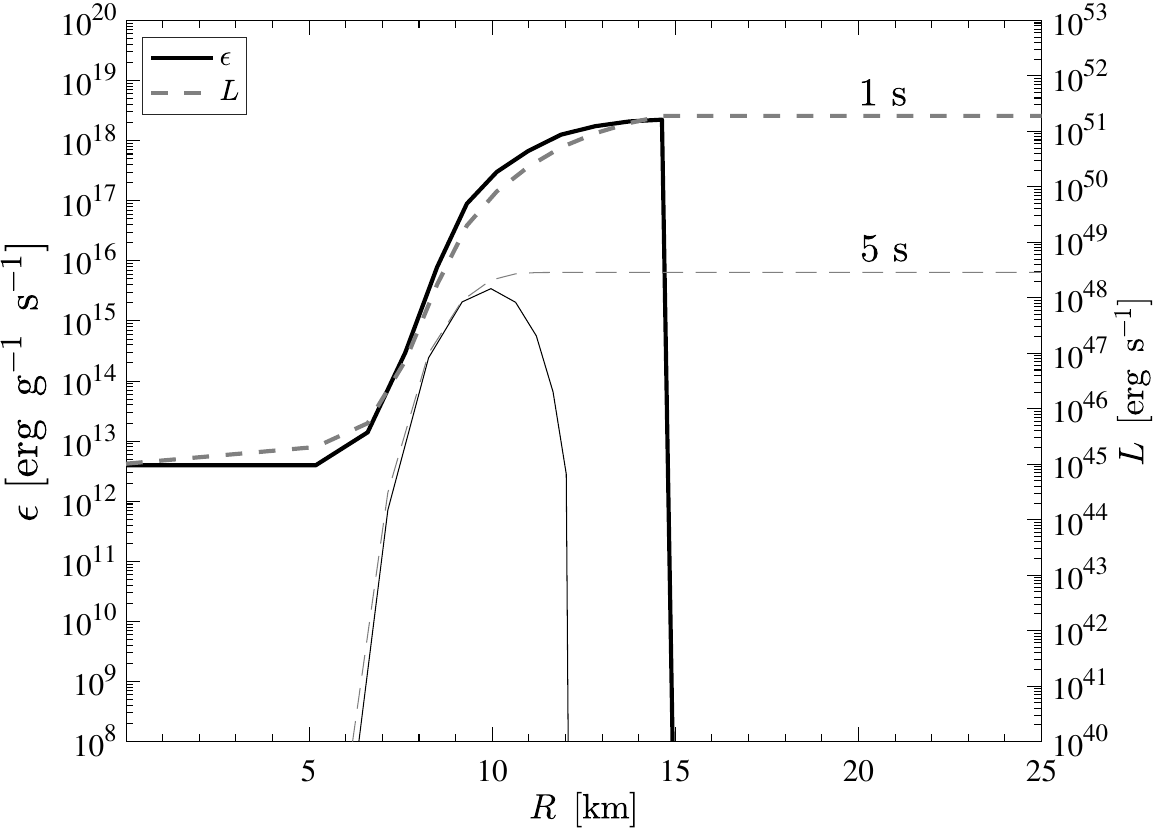}\label{fig:s25_dm}}
\caption{~Radial profiles of the dark particle emissivity $\epsilon$ (black solid lines on the left scale) and luminosity $L$ (gray dashed lines on the right scale) at post bounce times of  1~s and 5~s, corresponding to the HYPERONS$^*$ run, for which the conditions are shown in Fig.~\ref{fig:hydro}. \label{fig:rates}}
\end{center}
\end{figure}

At around 1~s (top panels of Fig.~\ref{fig:hydro}), the abundance of $\Lambda$ reach as high as 
about  $X_\Lambda\simeq 1\times 10^{-3}$ at the very center for the 18~M$_\odot$ model and 
about $X_\Lambda\simeq 5\times 10^{-3}$ for the 25~M$_\odot$ model. 
Going outwards in radius, the abundances rise, following the temperature profile, reaching their peak at around 
$X_\Lambda\simeq 5\times 10^{-3}$ for the 18~M$_\odot$ model and 
$X_\Lambda\simeq 2\times 10^{-2}$ for the 25~M$_\odot$ model. 
Figure~\ref{fig:rates} shows the corresponding dark particle emissivity (left scale) and luminosity (right scale) profiles. 
At around 1~s post bounce, dark sector emissivity and luminosity for the 18~M$_\odot$ model 
exceed those of the 25~M$_\odot$. 
Luminosities reach values of about $10^{52}$~erg for the former mode and about 
$5\times 10^{51}$~erg for the latter, which is also illustrated in the corresponding 
evolution of the dark sector luminosity in the top panels of Fig.~\ref{fig:lumin}. 
We attribute this to the larger branching ratio for the 18~M$_\odot$ model, despite having slightly
lower $\Lambda$ abundances and temperatures. 
Note also the sharp drop of the emissivities, after which the luminosities remain constant, which is due 
to the sudden drop of the $\Lambda$ abundance at around 13.75~km for the 18~M$_\odot$ model and 
at around 14.5~km for the 25~M$_\odot$ model, at 1~s post bounce. 
This is roughly to about one-half of saturation density, corresponding to conditions for the threshold 
for hyperons to exist within the HYPERON model EOS. 

The situation changes during the ongoing PNS deleptonization with the inclusion of the additional 
losses due to dark boson emission. The $X_0$ luminosity evolution is illustrated in Fig.~\ref{fig:lumin} 
(top panels), from where it becomes clear that the dark-boson losses dominate the cooling at around 
1~s post bounce, for both intermediate-mass and high-mass models. 
However, the excess cooling results in the rapid drop of the PNS temperature, significantly faster 
than in the reference HYPERONS simulations (see the bottom panels of Fig.~\ref{fig:hydro}). 
 Now, as the supernova interior features a lower temperature compared to the reference model 
without $X_0$ cooling, the feedback to the hyperon EOS is that the $\Lambda$ abundance is 
somewhat lower---note the strong temperature dependence of the hyperon EOS discussed in 
Sec.~\ref{sec:eos}---depending on the timescale during the PNS deleptonization. 
As a consequence, at some point during the later evolution, the $\Lambda$ abundance will be 
too low for the process $\Lambda \rightarrow n + X_0$ to still operate and the additional cooling 
due to $X_0$ emission ceases, featuring the continuous decrease of the $X_0$ luminosity; 
see the thick dashed lines in the top panels of Figs.~\ref{fig:s18_lumin} and \ref{fig:s25_lumin}. 
In a sense, the system self regulates it's excess cooling, however, depending sensitively on the BR.

This is illustrated for the 5~s post-bounce curves in Fig.~\ref{fig:rates} such that the $X_0$ 
luminosities in Fig.~\ref{fig:lumin} decrease below those of the total neutrinos and the later 
evolution will be determined by the neutrino losses. 
However, note that this is the case already at around 3~s post bounce for the intermediate-mass 
18~M$_\odot$ model, while only at around 7~s post bounce for the massive 25~M$_\odot$ model. 
It is related to the generally higher temperatures and central densities for the latter model, 
featuring generally higher abundances of hyperons. 

We note also the feedback from the excess cooling due to $X_0$ emission, namely the more 
rapid drop of the neutrino luminosities and average energies, for all flavours, during the PNS 
deleptonization phase that is dominated by $X_0$ cooling (see Fig.~\ref{fig:lumin}). 
We observe  the reduction of the neutrino emission timescale by a factor of roughly two for 
both intermediate-mass and high-mass models.

\section{Summary and conclusions}
\label{sec:summary}
Novel hyperonic EOS are studied in simulations of CCSN in spherical symmetry. 
These are based on general relativistic neutrino radiation hydrodynamics, featuring six-species 
Boltzmann neutrino transport. 
Particular emphasis has been put on the role of hyperons, which have long been studied in the 
context of cold neutron stars, where it has been speculated that the super-saturation density EOS 
softens due to the appearance of these additional and heavy degrees of freedom. 
In the context of CCSN studies, hyperons have so far only been taken into account in failed CCSN 
explosion studies that result in the formation of black holes 
\cite{Sumiyoshi2009ApJ690,Nakazato2012ApJ745,Bandyopadhyay2015ApJ809}, however, 
based on hadronic EOS with a limited set of hyperons. 
The focus of the present article is on the long-term evolution of neutrino-driven CCSN explosions, 
i.e. the deleptonization and later cooling phases of the nascent PNS, based on hadronic EOS 
including a comprehensive set of hyperons that include, $\Lambda$, $\Sigma^{0,\pm}$ and $\Xi^{0,\pm}$. 
The FSU2H$^*$ hyperon EOS employed in the present study is based on the RMF framework 
\cite{Kochankovski:2022rid,Kochankovski2024MNRAS528} with meson-nucleon coupling 
constants selected to reproduce nuclear matter and finite nuclei properties, to fulfil certain 
constraints of high dense matter coming from heavy-ion collisions as well as to be consistent 
with the observation of massive 2~M$_\odot$ pulsars. 

The results differ from those reported in the comprehensive CCSN hyperon EOS study of 
ref.~\cite{Hempel2014ApJS214} in various ways. 
The largest difference arises due to the implementation of larger sample of hadronic  states 
in the present paper. 
These modify the charge neutrality condition such that $Y_e+Y_\mu\neq Y_p$, which is 
the case in ref.~\cite{Hempel2014ApJS214} where no charged hyperons were considered. 
A second difference is related to the onset densities for the appearance of hyperons, which 
is the case at somewhat higher densities in ref.~\cite{Hempel2014ApJS214}, and the 
abundance of hyperons is significantly higher than for the present HYPERONS EOS. 
In particular, in ref.~\cite{Hempel2014ApJS214} the abundance of $\Lambda$ hyperons 
exceeds those of the neutrons at high baryon densities on the order of 
$\rho\simeq10^{15}~{\rm g}~{\rm cm}^{-3}$, even at low temperatures on the order of 
$1$--$10~{\rm MeV}$. 

In the CCSN simulations, the appearance of hyperons leaves a negligible impact 
on both the PNS structure during the entire early post-bounce evolution prior to the 
supernova shock revival and subsequent explosion onset. 
The reason is related to the generally low abundance of hyperons. 
The high-density EOS is dominated by the protons and neutrons. 
This situation changes only after several seconds, during the long-term PNS 
deleptonization evolution, when the central density rises continuously such that also the 
abundances of hyperons increase. 
However, even for massive PNS, on the order of about 2~M$_\odot$, the abundance 
of $\Lambda$ hyperons never exceeds more than few percent while the abundances 
of all heavier hyperons remain substantially lower, on the order of less than one percent. 
On the timescales considered here, on the order of several tens of seconds, the impact 
from the presence of hyperons on the PNS evolution remain negligible. CCSN observables, 
such as the neutrino signal, are hence insensitive to the composition at super-saturation 
density, explored in the present study  within a RMF model and neglecting possible 
contributions to the weak interactions. 
In particular, the latter assumption becomes invalid once neutrinos start to decouple
 in the hyperon phase. 
 This, however, has not been the case for the entire CCSN simulations under investigation. 

The implementation of the $\Lambda$ hyperon in the CCSN simulations enables us to 
study the impact of a possible dark sector. These arise from the flavour-violating transition of strange quarks into down quarks, taken 
into account here from the decay of $\Lambda$ hyperon into neutron via the emission 
of dark bosonic degrees of freedom. 
Other decays, such as that of $\Sigma^+$ are neglected due to their generally low 
abundances under the conditions of high isospin asymmetry encountered at the interior 
of the PNS.  

The excess cooling imposed by the inclusion of dark losses in the CCSN simulations 
results in a qualitative different evolution, in particular during the long-term PNS 
deleptonization phase when the abundance of $\Lambda$ hyperons reaches their maximum.
It is assumed that these massless particles escape the star once upon production. 
Possible re-scattering is omitted, which is consistent with the small couplings to the dark 
sector we are considering (c.f. ref.~\cite{Camalich:2020wac}). 
As a consequence, the PNS deleptonizes on a  shorter  timescale due to the different thermal evolution. 
As a feedback, similar as has been reported for the case of axion cooling (c.f. 
Refs.~\cite{Fischer2016PhRvD94_axions,Carenza:2019pxu,Fischer:2021jfm}) in the case of 
light axions from QCD processes, we find that the neutrino emission timescale is reduced 
by roughly a factor of two. 
This is consistent with the argument of ref.~\cite{Raffelt:1996wa}, confirming both values 
of the branching ratios, $10^{-8}$ for the low mass and $10^{-9}$ for the high mass models, 
as limiting cases corresponding to the intermediate-mass and high-mass PNS models of 
about 1.6~M$_\odot$ and 2.0~M$_\odot$, respectively. 
We explore these two cases at the level of neutrino-driven CCSN explosions of 
18~$M_\odot$ and 25~M$_\odot$ progenitors from the stellar evolution series of 
ref.~\cite{Woosley:2002zz}, which result in remnant PNS masses of about 1
.6~M$_\odot$ and 2.0~M$_\odot$ with lower central density and hence lower 
$\Lambda$ abundance for the former and the opposite for the latter models. 
We note here that the recent analysis at the level of the average $\bar\nu_e$-energy 
and the total $\bar\nu_e$-energy emitted, deduced from the KAM-II, IBM, BUST and 
LSD detections of SN1987A \cite{Fiorillo:2024PhRvD108,Li2024PhRvD109}, indicates 
that present standard core-collapse supernova models agree quantitatively with these 
measurements of the first few seconds of events while some tension remains with the 
late time events of Kam-II, IBM and BUST. 
The latter is due to generally short cooling timescales of all present long-term supernova 
simulations, including the PNS deleptonization and cooling phases. 
Note that considering the KAM-II data alone, current supernova models seemingly 
overestimate the detected $\bar\nu_e$ average energy, including the late-time detection. 
If these finding will be confirmed, in particular also from future supernova neutrino 
detection of the next galactic events, this would potentially put certain constraints on 
the requirement of non-standard additional cooling agents, such as the $X_0$ losses 
explored in this work due to the presence of strange hadronic degrees of freedom.

It has to be emphasised that the entire analysis, both comparing nucleonic and 
hyperonic EOS in CCSN simulations as well as the impact of associated dark sector 
cooling, is entirely model EOS dependent. 
The present analysis of the impact of dark photon losses in CCSN is dominated by the 
phase space of $\Lambda$ and the neutron degeneracy, which in turn are determined 
by the underlying hyperonic model. 
EOS with a quantitative different high-density behaviour of the strange degrees of freedom, 
e.g., featuring a lower(higher) abundance of $\Lambda$, will resemble also a lesser(stronger) 
impact on the excess dark cooling as is reported in the present article. 
All of these uncertainties can be related to the fact that the two-body and three-body interactions 
involving hyperons are poorly known. 
This is due to the fact that the available scattering data are still scarce and subject to quite 
large error bars, although promising results are becoming available from final-state interaction 
analyses and femtoscopy studies 
\cite{CLAS:2021gur,J-PARCE40:2021qxa,J-PARCE40:2021bgw,J-PARCE40:2022nvq,Fabbietti:2020bfg}. 
Also, data on hypernuclear structure can provide indirect information about the hyperon nuclear 
forces \cite{Feliciello:2015dua,Tamura:2013lwa,Gal:2016boi}. 
Moreover, advances can be anticipated on the experimental determination of two-body and 
three-body interactions involving hyperons and nucleons, e.g. at J-PARC, LHC or the future 
FAIR facility.

Another simplification applied here is related to the assumption of chemical equilibrium  
for strangeness  imposing equal chemical potentials, e.g., of neutrons and $\Lambda$ hyperons. 
On the other hand, the CCSN timescales, e.g., weak interactions, mass accretion and diffusion, 
might be potentially on the same order of magnitude as the timescale for hyperons to appear, 
however, through weak processes.
In other words, chemical equilibrium between strange and non-strange hadrons might not be 
fulfilled, similar as the condition of weak equilibrium is not fulfilled for muons for simulation times 
on the order of more than 30~s of the PNS deleptonization phase, as was shown in the Appendix 
of ref.~\cite{Fischer:2021jfm}. 
This requires the implementation of non-equilibrium methods in order to solve for the problem 
of strangeness in CCSN, which we leave for future explorations.

\section*{Acknowledgments}
We would like to thank Pok~Man~Lo for helpful comments and inspiring discussions.
TF acknowledges support from the Polish National Science Centre (NCN) under grant number 2023/49/B/ST9/03941. 
JMC thanks MICINN for funding through the grant ``DarkMaps'' PID2022-142142NB-I00 and from the European Union through the grant ``UNDARK'' of the Widening participation and spreading excellence programme (project number 101159929).   
HK acknowledges support from the PRE2020-093558 Doctoral Grant of the spanish MCIN/AEI/10.13039/ 501100011033/. 
LT acknowledges support from the project CEX2020-001058-M (Unidades de Excelencia ``Mar\'{\i}a de Maeztu") and PID2022-139427NB-I00, financed by the Spanish MCIN/ AEI/10.13039/501100011033/, as well as by the EU STRONG-2020 project, under the program H2020-INFRAIA-2018-1 grant agreement no. 824093, from the DFG through projects No. 315477589 - TRR 211 (Strong-interaction matter under extreme conditions), and from the Generalitat de Catalunya under contract 2021 SGR 171.
All scientific computations were performed at the Wroclaw Centre for Scientific Computing and Networking (WCSS) under 'grant328'.

%


\bibliographystyle{JHEP}

\providecommand{\href}[2]{#2}\begingroup\raggedright\begin{thebibliography}{}

\end{thebibliography}\endgroup


\begin{thebibliography}{100}

\bibitem{Janka07}
H.-Th.~{Janka}, K.~{Langanke}, A.~{Marek}, G.~{Mart{\'{\i}}nez-Pinedo} and
  B.~{M{\"u}ller}, \emph{{Theory of core-collapse supernovae}},
  \href{https://doi.org/10.1016/j.physrep.2007.02.002}{\emph{\physrep}
  {\bfseries 442} (2007) 38}
  [\href{https://arxiv.org/abs/arXiv:astro-ph/0612072}{
  {\ttfamily arXiv:0612072}}].

\bibitem{mirizzi16}
A.~{Mirizzi}, I.~{Tamborra}, H.-Th.~{Janka}, N.~{Saviano}, K.~{Scholberg},
  R.~{Bollig} et~al., \emph{{Supernova neutrinos: production, oscillations and
  detection}}, \href{https://doi.org/10.1393/ncr/i2016-10120-8}{\emph{Nuovo
  Cimento Rivista Serie} {\bfseries 39} (2016) 1}
  [\href{https://arxiv.org/abs/1508.00785}{
  {\ttfamily arXiv:1508.00785}}].

\bibitem{Burrows2021Natur589}
A.~{Burrows} and D.~{Vartanyan}, \emph{{Core-collapse supernova explosion
  theory}}, \href{https://doi.org/10.1038/s41586-020-03059-w}{\emph{\nat}
  {\bfseries 589} (2021) 29}
  [\href{https://arxiv.org/abs/2009.14157}{{\ttfamily arXiv:2009.14157}}].

\bibitem{LeBlancWilson70}
J.~M.~{LeBlanc} and J.~R.~{Wilson}, \emph{{A Numerical Example of the Collapse of
  a Rotating Magnetized Star}},
  \href{https://doi.org/10.1086/150558}{\emph{\apj} {\bfseries 161} (1970) 541}.

\bibitem{Bethe85}
H.~A.~{Bethe} and J.~R.~{Wilson}, \emph{{Revival of a stalled supernova shock by
  neutrino heating}}, \href{https://doi.org/10.1086/163343}{\emph{\apj}
  {\bfseries 295} (1985) 14}.

\bibitem{kitaura}
F.~S.~{Kitaura}, H.-Th.~{Janka} and W.~{Hillebrandt}, \emph{{Explosions of
  O-Ne-Mg cores, the Crab supernova, and subluminous type II-P supernovae}},
  \href{https://doi.org/10.1051/0004-6361:20054703}{\emph{\aap} {\bfseries 450}
  (2006) 345} 
  [\href{https://arxiv.org/abs/arXiv:astro-ph/0512065}{{\ttfamily arXiv:0512065}}].

\bibitem{Fischer10}
T.~{Fischer}, S.C.~{Whitehouse}, A.~{Mezzacappa}, F.~K.~{Thielemann} and
  M.~{Liebend{\"o}rfer}, \emph{{Protoneutron star evolution and the
  neutrino-driven wind in general relativistic neutrino radiation hydrodynamics
  simulations}}, \href{https://doi.org/10.1051/0004-6361/200913106}{\emph{\aap}
  {\bfseries 517} (2010) A80}
  [\href{https://arxiv.org/abs/0908.1871}{{\ttfamily arXiv:0908.1871}}].

\bibitem{Huedepohl10}
L.~{H{\"u}depohl}, B.~{M{\"u}ller}, H.-Th.~{Janka}, A.~{Marek} and
  G.G.~{Raffelt}, \emph{{Neutrino Signal of Electron-Capture Supernovae from
  Core Collapse to Cooling}},
  \href{https://doi.org/10.1103/PhysRevLett.104.251101}{\emph{\prl} {\bfseries
  104} (2010) 251101} 
  [\href{https://arxiv.org/abs/0912.0260}{{\ttfamily arXiv:0912.0260}}].

\bibitem{Tauris15}
T.~M.~{Tauris}, N.~{Langer} and P.~{Podsiadlowski}, \emph{{Ultra-stripped
  supernovae: progenitors and fate}},
  \href{https://doi.org/10.1093/mnras/stv990}{\emph{\mnras} {\bfseries 451}
  (2015) 2123} 
  [\href{https://arxiv.org/abs/1505.00270}{{\ttfamily arXiv:1505.00270}}].

\bibitem{De:2018Sci362}
K.~{De}, M.~M.~{Kasliwal}, E.~O.~{Ofek}, T.~J.~{Moriya}, J.~{Burke}, Y.~{Cao}
  et~al., \emph{{A hot and fast ultra-stripped supernova that likely formed a
  compact neutron star binary}},
  \href{https://doi.org/10.1126/science.aas8693}{\emph{Science} {\bfseries 362}
  (2018) 201} [\href{https://arxiv.org/abs/1810.05181}{{\ttfamily
  arXiv:1810.05181}}].

\bibitem{BMuller18}
B.~{M{\"u}ller}, D.~W.~{Gay}, A.~{Heger}, T.~M.~{Tauris} and S.~A.~{Sim},
  \emph{{Multidimensional simulations of ultrastripped supernovae to shock
  breakout}}, \href{https://doi.org/10.1093/mnras/sty1683}{\emph{\mnras}
  {\bfseries 479} (2018) 3675}
  [\href{https://arxiv.org/abs/1803.03388}{{\ttfamily arXiv:1803.03388}}].

\bibitem{BMuller19}
B.~{M{\"u}ller}, T.~M.~{Tauris}, A.~{Heger}, P.~{Banerjee}, Y.-Z.~{Qian},
  J.~{Powell} et~al., \emph{{Three-dimensional simulations of neutrino-driven
  core-collapse supernovae from low-mass single and binary star progenitors}},
  \href{https://doi.org/10.1093/mnras/stz216}{\emph{\mnras} {\bfseries 484}
  (2019) 3307} [\href{https://arxiv.org/abs/1811.05483}{{\ttfamily
  arXiv:1811.05483}}].

\bibitem{Ertl:2020ApJ890}
T.~{Ertl}, S.E.~{Woosley}, T.~{Sukhbold} and H.-Th.~{Janka}, \emph{{The Explosion
  of Helium Stars Evolved with Mass Loss}},
  \href{https://doi.org/10.3847/1538-4357/ab6458}{\emph{\apj} {\bfseries 890}
  (2020) 51} [\href{https://arxiv.org/abs/1910.01641}{{\ttfamily arXiv:1910.01641}}].

\bibitem{Stockinger:2020MNRAS496}
G.~{Stockinger}, H.-Th.~{Janka}, D.~{Kresse}, T.~{Melson}, T.~{Ertl}, M.~{Gabler}
  et~al., \emph{{Three-dimensional models of core-collapse supernovae from
  low-mass progenitors with implications for Crab}},
  \href{https://doi.org/10.1093/mnras/staa1691}{\emph{\mnras} {\bfseries 496}
  (2020) 2039} [\href{https://arxiv.org/abs/2005.02420}{{\ttfamily
  arXiv:2005.02420}}].

\bibitem{Wang24}
T.~{Wang} and A.~{Burrows}, \emph{{Supernova Explosions of the Lowest-mass
  Massive Star Progenitors}},
  \href{https://doi.org/10.3847/1538-4357/ad5009}{\emph{\apj} {\bfseries 969}
  (2024) 74} [\href{https://arxiv.org/abs/2405.06024}{{\ttfamily arXiv:2405.06024}}].

\bibitem{Sagert09}
I.~{Sagert}, T.~{Fischer}, M.~{Hempel}, G.~{Pagliara}, J.~{Schaffner-Bielich},
  A.~{Mezzacappa} et~al., \emph{{Signals of the QCD Phase Transition in
  Core-Collapse Supernovae}},
  \href{https://doi.org/10.1103/PhysRevLett.102.081101}{\emph{\prl} {\bfseries
  102} (2009) 081101} 
  [\href{https://arxiv.org/abs/0809.4225}{{\ttfamily arXiv:0809.4225}}].

\bibitem{Fischer18}
T.~{Fischer}, N.-U.~F.~{Bastian}, M.-R.~{Wu}, P.~{Baklanov}, E.~{Sorokina},
  S.~{Blinnikov} et~al., \emph{{Quark deconfinement as a supernova explosion
  engine for massive blue supergiant stars}},
  \href{https://doi.org/10.1038/s41550-018-0583-0}{\emph{Nature Astronomy}
  {\bfseries 2} (2018) 980} 
  [\href{https://arxiv.org/abs/1712.08788}{{\ttfamily arXiv:1712.08788}}].

\bibitem{Zha20}
S.~{Zha}, E.~P.~{O'Connor}, M.-c.~{Chu}, L.-M.~{Lin} and S.M.~{Couch},
  \emph{{Gravitational-wave Signature of a First-order Quantum Chromodynamics
  Phase Transition in Core-Collapse Supernovae}},
  \href{https://doi.org/10.1103/PhysRevLett.125.051102}{\emph{\prl} {\bfseries
  125} (2020) 051102}
  [\href{https://arxiv.org/abs/2007.04716}{{\ttfamily arXiv:2007.04716}}].

\bibitem{Fischer:2021}
T.~{Fischer}, \emph{{QCD phase transition drives supernova explosion of a very
  massive star}},
  \href{https://doi.org/10.1140/epja/s10050-021-00571-z}{\emph{Eur.\ Phys.\ J.\
  A} {\bfseries 57} (2021) 270}
  [\href{https://arxiv.org/abs/2108.00196}{{\ttfamily arXiv:2108.00196}}].

\bibitem{Kuroda2022}
T.~{Kuroda}, T.~{Fischer}, T.~{Takiwaki} and K.~{Kotake}, \emph{{Core-collapse
  Supernova Simulations and the Formation of Neutron Stars, Hybrid Stars, and
  Black Holes}}, \href{https://doi.org/10.3847/1538-4357/ac31a8}{\emph{\apj}
  {\bfseries 924} (2022) 38}
  [\href{https://arxiv.org/abs/2109.01508}{{\ttfamily arXiv:2109.01508}}].

\bibitem{KhosraviLargani2024ApJ964}
N.~{Khosravi Largani}, T.~{Fischer} and N.-U.~F.~{Bastian}, \emph{{Constraining
  the Onset Density for the QCD Phase Transition with the Neutrino Signal from
  Core-collapse Supernovae}},
  \href{https://doi.org/10.3847/1538-4357/ad24f2}{\emph{\apj} {\bfseries 964}
  (2024) 143} 
  [\href{https://arxiv.org/abs/2304.12316}{{\ttfamily arXiv:2304.12316}}].

\bibitem{KurodaShibata:2023PhRvD107}
T.~{Kuroda} and M.~{Shibata}, 
\emph{{Spontaneous scalarization as a new core-collapse supernova mechanism and its multimessenger signals}},
\href{https://10.1103/PhysRevD.107.103025}
{\emph{\prd} {\bfseries 107} (2023) 103025} 
[\href{https://arxiv.org/abs/2302.09853}{{\ttfamily arXiv:2302.09853}}].

\bibitem{Fischer17}
T.~{Fischer}, N.-U.~F.~{Bastian}, D.~{Blaschke}, M.~{Cierniak}, M.~{Hempel},
  T.~{Kl{\"a}hn} et~al., \emph{{The State of Matter in Simulations of
  Core-Collapse supernovae{\textemdash}Reflections and Recent Developments}},
  \href{https://doi.org/10.1017/pasa.2017.63}{\emph{\pasa} {\bfseries 34}
  (2017) e067} 
  [\href{https://arxiv.org/abs/1711.07411}{{\ttfamily arXiv:1711.07411}}].

\bibitem{Bauswein19}
A.~{Bauswein}, N.-U.~F.~{Bastian}, D.B.~{Blaschke}, K.~{Chatziioannou},
  J.A~.~{Clark}, T.~{Fischer} et~al., \emph{{Identifying a First-Order Phase
  Transition in Neutron-Star Mergers through Gravitational Waves}},
  \href{https://doi.org/10.1103/PhysRevLett.122.061102}{\emph{\prl} {\bfseries
  122} (2019) 061102}
  [\href{https://arxiv.org/abs/1809.01116}{{\ttfamily arXiv:1809.01116}}].

\bibitem{Most19}
E.~R.~{Most}, L.~J.~{Papenfort}, V.~{Dexheimer}, M.~{Hanauske}, S.~{Schramm},
  H.~{St{\"o}cker} et~al., \emph{{Signatures of Quark-Hadron Phase Transitions
  in General-Relativistic Neutron-Star Mergers}},
  \href{https://doi.org/10.1103/PhysRevLett.122.061101}{\emph{\prl} {\bfseries
  122} (2019) 061101} 
  [\href{https://arxiv.org/abs/1807.03684}{{\ttfamily arXiv:1807.03684}}].

\bibitem{Sumiyoshi2009ApJ690}
K.~{Sumiyoshi}, C.~{Ishizuka}, A.~{Ohnishi}, S.~{Yamada} and H.~{Suzuki},
  \emph{{Emergence of Hyperons in Failed Supernovae: Trigger of the Black Hole
  Formation}}, \href{https://doi.org/10.1088/0004-637X/690/1/L43}{\emph{\apjl}
  {\bfseries 690} (2009) L43}
  [\href{https://arxiv.org/abs/0811.4237}{{\ttfamily arXiv:0811.4237}}].

\bibitem{Fischer09}
T.~{Fischer}, S.~C.~{Whitehouse}, A.~{Mezzacappa}, F.-K.~{Thielemann} and
  M.~{Liebend{\"o}rfer}, \emph{{The neutrino signal from protoneutron star
  accretion and black hole formation}},
  \href{https://doi.org/10.1051/0004-6361/200811055}{\emph{\aap} {\bfseries
  499} (2009) 1} [\href{https://arxiv.org/abs/0809.5129}{{\ttfamily
  arXiv:0809.5129}}].

\bibitem{O'Connor11}
E.~{O'Connor} and C.D.~{Ott}, \emph{{Black Hole Formation in Failing
  Core-Collapse Supernovae}},
  \href{https://doi.org/10.1088/0004-637X/730/2/70}{\emph{\apj} {\bfseries 730}
  (2011) 70} [\href{https://arxiv.org/abs/1010.5550}{{\ttfamily arXiv:1010.5550}}].

\bibitem{Nakazato2012ApJ745}
K.~{Nakazato}, S.~{Furusawa}, K.~{Sumiyoshi}, A.~{Ohnishi}, S.~{Yamada} and
  H.~{Suzuki}, \emph{{Hyperon Matter and Black Hole Formation in Failed
  Supernovae}}, \href{https://doi.org/10.1088/0004-637X/745/2/197}{\emph{\apj}
  {\bfseries 745} (2012) 197}
  [\href{https://arxiv.org/abs/1111.2900}{{\ttfamily arXiv:1111.2900}}].

\bibitem{Bandyopadhyay2015ApJ809}
P.~{Char}, S.~{Banik} and D.~{Bandyopadhyay}, \emph{{A Comparative Study of
  Hyperon Equations of State in Supernova Simulations}},
  \href{https://doi.org/10.1088/0004-637X/809/2/116}{\emph{\apj} {\bfseries
  809} (2015) 116} 
  [\href{https://arxiv.org/abs/1508.01854}{{\ttfamily arXiv:1508.01854}}].

\bibitem{FiorellaBurgio2018NuclearSupernovae}
G.F.~Burgio and A.F.~Fantina, \emph{Nuclear equation of state for compact stars
  and supernovae},  in \emph{The Physics and Astrophysics of Neutron Stars},
  pp.~255--335, Springer International Publishing (2018).

\bibitem{Tolos:2020aln}
L.~Tolos and L.~Fabbietti, \emph{{Strangeness in Nuclei and Neutron Stars}},
  \href{https://doi.org/10.1016/j.ppnp.2020.103770}{\emph{Prog. Part. Nucl.
  Phys.} {\bfseries 112} (2020) 103770}
  [\href{https://arxiv.org/abs/2002.09223}{{\ttfamily arXiv:2002.09223}}].

\bibitem{Burgio:2021vgk}
G.F.~Burgio, H.J.~Schulze, I.~Vidana and J.B.~Wei, \emph{{Neutron stars and the
  nuclear equation of state}},
  \href{https://doi.org/10.1016/j.ppnp.2021.103879}{\emph{Prog. Part. Nucl.
  Phys.} {\bfseries 120} (2021) 103879}
  [\href{https://arxiv.org/abs/2105.03747}{{\ttfamily arXiv:2105.03747}}].

\bibitem{MUSES:2023hyz}
{\scshape MUSES} collaboration, \emph{{Theoretical and experimental constraints
  for the equation of state of dense and hot matter}},
  \href{https://doi.org/10.1007/s41114-024-00049-6}{\emph{Living Rev. Rel.}
  {\bfseries 27} (2024) 3} 
  [\href{https://arxiv.org/abs/2303.17021}{{\ttfamily arXiv:2303.17021}}].

\bibitem{Demorest2010ShapiroStar}
P.~Demorest, T.~Pennucci, S.~Ransom, M.~Roberts and J.~Hessels, \emph{{Shapiro
  delay measurement of a two solar mass neutron star}},
  \href{https://doi.org/10.1038/nature09466}{\emph{Nature} {\bfseries 467}
  (2010) 1081}.
  [\href{https://arxiv.org/abs/1010.5788}{{\ttfamily arXiv:1010.5788}}].

\bibitem{Antoniadis13}
J.~{Antoniadis}, P.C.C.~{Freire}, N.~{Wex}, T.M.~{Tauris}, R.S.~{Lynch},
  M.H.~{van Kerkwijk} et~al., \emph{{A Massive Pulsar in a Compact Relativistic
  Binary}}, \href{https://doi.org/10.1126/science.1233232}{\emph{Science}
  {\bfseries 340} (2013) 448}
  [\href{https://arxiv.org/abs/1304.6875}{{\ttfamily arXiv:1304.6875}}].

\bibitem{Fonseca:2021}
E.~{Fonseca}, H.T.~{Cromartie}, T.T.~{Pennucci}, P.S.~{Ray}, A.Y.~{Kirichenko},
  S.M.~{Ransom} et~al., \emph{{Refined Mass and Geometric Measurements of the
  High-mass PSR J0740+6620}},
  \href{https://doi.org/10.3847/2041-8213/ac03b8}{\emph{\apjl} {\bfseries 915}
  (2021) L12} 
  [\href{https://arxiv.org/abs/2104.00880}{{\ttfamily arXiv:2104.00880}}].

\bibitem{Cromartie2020RelativisticPulsar}
{\scshape NANOGrav} collaboration, \emph{{Relativistic Shapiro delay
  measurements of an extremely massive millisecond pulsar}},
  \href{https://doi.org/10.1038/s41550-019-0880-2}{\emph{Nature Astron.}
  {\bfseries 4} (2019) 72} 
  [\href{https://arxiv.org/abs/1904.06759}{{\ttfamily arXiv:1904.06759}}].

\bibitem{Romani:2022jhd}
R.W.~Romani, D.~Kandel, A.V.~Filippenko, T.G.~Brink and W.~Zheng, \emph{{PSR
  J0952\ensuremath{-}0607: The Fastest and Heaviest Known Galactic Neutron
  Star}}, \href{https://doi.org/10.3847/2041-8213/ac8007}{\emph{Astrophys. J.
  Lett.} {\bfseries 934} (2022) L17}
  [\href{https://arxiv.org/abs/2207.05124}{{\ttfamily arXiv:2207.05124}}].

\bibitem{Petschauer:2020urh}
S.~Petschauer, J.~Haidenbauer, N.~Kaiser, U.-G.~Mei\ss{}ner and W.~Weise,
  \emph{{Hyperon-nuclear interactions from SU(3) chiral effective field
  theory}}, \href{https://doi.org/10.3389/fphy.2020.00012}{\emph{Front. in
  Phys.} {\bfseries 8} (2020) 12}
  [\href{https://arxiv.org/abs/2002.00424}{{\ttfamily arXiv:2002.00424}}].

\bibitem{Pok:2020PhRvD102}
P.~M.~{Lo}, \emph{{Density of states of a coupled-channel system}},
  \href{https://doi.org/10.1103/PhysRevD.102.034038}{\emph{\prd} {\bfseries
  102} (2020) 034038} 
  [\href{https://arxiv.org/abs/2007.03392}{{\ttfamily arXiv:2007.03392}}].

\bibitem{Pok:2021PhRvC103}
J.~{Cleymans}, P.~M.~{Lo}, K.~{Redlich} and N.~{Sharma}, \emph{{Multiplicity
  dependence of (multi)strange baryons in the canonical ensemble with phase
  shift corrections}},
  \href{https://doi.org/10.1103/PhysRevC.103.014904}{\emph{\prc} {\bfseries
  103} (2021) 014904} 
  [\href{https://arxiv.org/abs/2009.04844}{{\ttfamily arXiv:2009.04844}}].

\bibitem{Pok:2021EPJA57}
P.~M.~{Lo}, \emph{{Thermal study of a coupled-channel system: a brief review}},
  \href{https://doi.org/10.1140/epja/s10050-021-00378-y}{\emph{European
  Physical Journal A} {\bfseries 57} (2021) 60}.

\bibitem{CLAS:2021gur}
{\scshape CLAS} collaboration, \emph{{Improved $\Lambda p$ Elastic Scattering
  Cross Sections Between 0.9 and 2.0 GeV/c and Connections to the Neutron Star
  Equation of State}},
  \href{https://doi.org/10.1103/PhysRevLett.127.272303}{\emph{Phys. Rev. Lett.}
  {\bfseries 127} (2021) 272303}
  [\href{https://arxiv.org/abs/2108.03134}{{\ttfamily arXiv:2108.03134}}].

\bibitem{J-PARCE40:2021qxa}
{\scshape J-PARC E40} collaboration, \emph{{Measurement of the differential
  cross sections of the $\Sigma^-p$ elastic scattering in momentum range 470 to
  850 MeV/c}}, \href{https://doi.org/10.1103/PhysRevC.104.045204}{\emph{Phys.
  Rev. C} {\bfseries 104} (2021) 045204}
  [\href{https://arxiv.org/abs/2104.13608}{{\ttfamily arXiv:2104.13608}}].

\bibitem{J-PARCE40:2021bgw}
{\scshape J-PARC E40} collaboration, \emph{{Precise measurement of differential
  cross sections of the $\Sigma^-p \to \Lambda n$ reaction in momentum range
  470-650 MeV$/c$}},
  \href{https://doi.org/10.1103/PhysRevLett.128.072501}{\emph{Phys. Rev. Lett.}
  {\bfseries 128} (2022) 072501}
  [\href{https://arxiv.org/abs/2111.14277}{{\ttfamily arXiv:2111.14277}}].

\bibitem{J-PARCE40:2022nvq}
{\scshape J-PARC E40} collaboration, \emph{{Measurement of differential cross
  sections~for \ensuremath{\Sigma}+p elastic scattering in the momentum range
  0.44\textendash{}0.80\,GeV/c}},
  \href{https://doi.org/10.1093/ptep/ptac101}{\emph{PTEP} {\bfseries 2022}
  (2022) 093D01}
  [\href{https://arxiv.org/abs/2203.08393}{{\ttfamily arXiv:2203.08393}}].

\bibitem{Fabbietti:2020bfg}
L.~Fabbietti, V.~Mantovani~Sarti and O.~Vazquez~Doce, \emph{{Study of the
  Strong Interaction Among Hadrons with Correlations at the LHC}},
  \href{https://doi.org/10.1146/annurev-nucl-102419-034438}
  {\emph{Ann. Rev. Nucl. Part. Sci.} {\bfseries 71} (2021) 377}
  [\href{https://arxiv.org/abs/2012.09806}{{\ttfamily arXiv:2012.09806}}].

\bibitem{Feliciello:2015dua}
A.~Feliciello and T.~Nagae, \emph{{Experimental review of hypernuclear physics:
  recent achievements and future perspectives}},
  \href{https://doi.org/10.1088/0034-4885/78/9/096301}
  {\emph{Rept. Prog. Phys.} {\bfseries 78} (2015) 096301}.

\bibitem{Tamura:2013lwa}
H.~{Tamura} et~al., \emph{{Gamma-ray spectroscopy of hypernuclei - present and
  future}}, \href{https://doi.org/10.1016/j.nuclphysa.2013.03.014}
  {\emph{Nucl. Phys. A} {\bfseries 914} (2013) 99}.

\bibitem{Gal:2016boi}
A.~{Gal}, E.V.~{Hungerford} and D.J.~{Millener}, \emph{{Strangeness in nuclear
  physics}}, \href{https://doi.org/10.1103/RevModPhys.88.035004}{\emph{Rev.
  Mod. Phys.} {\bfseries 88} (2016) 035004}
  [\href{https://arxiv.org/abs/1605.00557}{{\ttfamily arXiv:1605.00557}}].

\bibitem{Raduta:2021coc}
A.~R.~Raduta, F.~Nacu and M.~Oertel, 
  \emph{{Equations of state for hot neutron stars}}, 
  \href{https://doi.org/10.1140/epja/s10050-021-00628-z}
  {\emph{Eur. Phys. J. A} {\bfseries 57} (2021) 329}
  [\href{https://arxiv.org/abs/2109.00251}{{\ttfamily arXiv:2109.00251}}].

\bibitem{Raduta:2022elz}
A.~R.~Raduta, \emph{{Equations of state for hot neutron stars-II. The role of
  exotic particle degrees of freedom}},
  \href{https://doi.org/10.1140/epja/s10050-022-00772-0}
  {\emph{Eur. Phys. J. A} {\bfseries 58} (2022) 115}
  [\href{https://arxiv.org/abs/2205.03177}{{\ttfamily arXiv:2205.03177}}].

\bibitem{Oertel:2016bki}
M.~Oertel, M.~Hempel, T.~Kl\"ahn and S.~Typel, 
  \emph{{Equations of state for supernovae and compact stars}},
  \href{https://doi.org/10.1103/RevModPhys.89.015007}
  {\emph{Rev. Mod. Phys.} {\bfseries 89} (2017) 015007}
  [\href{https://arxiv.org/abs/1610.03361}{{\ttfamily arXiv:1610.03361}}].

\bibitem{Dexheimer:2022qhn}
V.~Dexheimer, M.~Mancini, M.~Oertel, C.~Provid\^encia, L.~Tolos and S.~Typel,
  \emph{{Quick Guides for Use of the CompOSE Data Base}},
  \href{https://doi.org/10.3390/particles5030028}{\emph{Particles} {\bfseries
  5} (2022) 346} 
  [\href{https://arxiv.org/abs/2311.04715}{{\ttfamily arXiv:2311.04715}}].

\bibitem{CompOSECoreTeam:2022ddl}
{\scshape CompOSE Core Team} collaboration,  
  \emph{{CompOSE Reference Manual}}, 
  \href{https://doi.org/10.1140/epja/s10050-022-00847-y}
  {\emph{Eur. Phys. J. A} {\bfseries 58} (2022) 221}
  [\href{https://arxiv.org/abs/2203.03209}{{\ttfamily arXiv:2203.03209}}].

\bibitem{Raffelt:1987yt}
G.~G.~{Raffelt} and D.~{Seckel}, \emph{{Bounds on Exotic Particle Interactions
  from SN 1987a}},
  \href{https://doi.org/10.1103/PhysRevLett.60.1793}{\emph{Phys. Rev. Lett.}
  {\bfseries 60} (1988) 1793}.

\bibitem{Raffelt:1996wa}
G.~G.~{Raffelt}, \emph{{Stars as laboratories for fundamental physics : the
  astrophysics of neutrinos, axions, and other weakly interacting particles}}
  (1996).

\bibitem{Turner:1987by}
M.~S.~Turner, \emph{{Axions from SN 1987a}},
  \href{https://doi.org/10.1103/PhysRevLett.60.1797}
  {\emph{Phys. Rev. Lett.} {\bfseries 60} (1988) 1797}.

\bibitem{Mayle:1987as}
R.~Mayle, J.R.~Wilson, J.R.~Ellis, K.A.~Olive, D.N.~Schramm and G.~Steigman,
  \emph{{Constraints on Axions from SN 1987a}},
  \href{https://doi.org/10.1016/0370-2693(88)91595-X}{\emph{Phys. Lett. B}
  {\bfseries 203} (1988) 188}.

\bibitem{Burrows:1988ah}
A.~Burrows, M.~S.~Turner and R.~Brinkmann,
  \emph{{Axions and SN 1987a}},
  \href{https://doi.org/10.1103/PhysRevD.39.1020}{\emph{Phys. Rev. D}
  {\bfseries 39} (1989) 1020}.

\bibitem{Burrows:1990pk}
A.~Burrows, M.~Ressell and M.~S.~Turner, 
  \emph{{Axions and SN1987A: Axion trapping}}, 
  \href{https://doi.org/10.1103/PhysRevD.42.3297}
  {\emph{Phys. Rev. D} {\bfseries 42} (1990) 3297}.

\bibitem{Carenza:2019pxu}
P.~Carenza, T.~Fischer, M.~Giannotti, G.~Guo, G.~Mart\'\i{}nez-Pinedo and A.~Mirizzi, 
  \emph{{Improved axion emissivity from a supernova via nucleon-nucleon bremsstrahlung}},
  \href{https://doi.org/10.1088/1475-7516/2019/10/016}
  {\emph{JCAP} {\bfseries 10} (2019) 016} 
  [\href{https://arxiv.org/abs/1906.11844}{{\ttfamily arXiv:1906.11844}}].

\bibitem{Choi:2021ign}
K.~Choi, H.~J.~Kim, H.~Seong and C.~S.~Shin, 
  \emph{{Axion emission from supernova with axion-pion-nucleon contact interaction}},
  \href{https://doi.org/10.1007/JHEP02(2022)143}{\emph{JHEP} {\bfseries 02} (2022) 143} 
  [\href{https://arxiv.org/abs/2110.01972}{{\ttfamily arXiv:2110.01972}}].

\bibitem{DiLuzio:2020wdo}
L.~Di~Luzio, M.~Giannotti, E.~Nardi and L.~Visinelli, 
  \emph{{The landscape of QCD axion models}},
  \href{https://doi.org/10.1016/j.physrep.2020.06.002}
  {\emph{Phys. Rept.} {\bfseries 870} (2020) 1} 
  [\href{https://arxiv.org/abs/2003.01100}{{\ttfamily arXiv:2003.01100}}].

\bibitem{Carenza:2020cis}
P.~Carenza, B.~Fore, M.~Giannotti, A.~Mirizzi and S.~Reddy, 
  \emph{{Enhanced Supernova Axion Emission and its Implications}},
  \href{https://doi.org/10.1103/PhysRevLett.126.071102}
  {\emph{Phys. Rev. Lett.}{\bfseries 126} (2021) 071102}
  [\href{https://arxiv.org/abs/2010.02943}{{\ttfamily arXiv:2010.02943}}].

\bibitem{Fischer:2021jfm}
T.~Fischer, P.~Carenza, B.~Fore, M.~Giannotti, A.~Mirizzi and S.~Reddy,
  \emph{{Observable signatures of enhanced axion emission from protoneutron
  stars}}, \href{https://doi.org/10.1103/PhysRevD.104.103012}{\emph{Phys. Rev.
  D} {\bfseries 104} (2021) 103012}
  [\href{https://arxiv.org/abs/2108.13726}{{\ttfamily arXiv:2108.13726}}].

\bibitem{Rrapaj:2015wgs}
E.~Rrapaj and S.~Reddy, \emph{{Nucleon-nucleon bremsstrahlung of dark gauge
  bosons and revised supernova constraints}},
  \href{https://doi.org/10.1103/PhysRevC.94.045805}{\emph{Phys. Rev. C}
  {\bfseries 94} (2016) 045805}
  [\href{https://arxiv.org/abs/1511.09136}{{\ttfamily arXiv:1511.09136}}].

\bibitem{Chang:2018rso}
J.~H.~Chang, R.~Essig and S.~D.~McDermott, \emph{{Supernova 1987A Constraints on
  Sub-GeV Dark Sectors, Millicharged Particles, the QCD Axion, and an
  Axion-like Particle}},
  \href{https://doi.org/10.1007/JHEP09(2018)051}{\emph{JHEP} {\bfseries 09}
  (2018) 051} 
  [\href{https://arxiv.org/abs/1803.00993}{{\ttfamily arXiv:1803.00993}}].

\bibitem{Calore:2021klc}
F.~Calore, P.~Carenza, M.~Giannotti, J.~Jaeckel, G.~Lucente and A.~Mirizzi,
  \emph{{Supernova bounds on axionlike particles coupled with nucleons and
  electrons}}, \href{https://doi.org/10.1103/PhysRevD.104.043016}{\emph{Phys.
  Rev. D} {\bfseries 104} (2021) 043016}
  [\href{https://arxiv.org/abs/2107.02186}{{\ttfamily arXiv:2107.02186}}].

\bibitem{Balaji:2022noj}
S.~{Balaji}, P.S.B.~{Dev}, J.~{Silk} and Y.~{Zhang}, \emph{{Improved stellar
  limits on a light CP-even scalar}},
  \href{https://doi.org/10.1088/1475-7516/2022/12/024}{\emph{\jcap} {\bfseries
  2022} (2022) 024} 
  [\href{https://arxiv.org/abs/2205.01669}{{\ttfamily arXiv:2205.01669}}].

\bibitem{Lella:2022uwi}
A.~{Lella}, P.~{Carenza}, G.~{Lucente}, M.~{Giannotti} and A.~{Mirizzi},
  \emph{{Protoneutron stars as cosmic factories for massive axionlike
  particles}}, \href{https://doi.org/10.1103/PhysRevD.107.103017}{\emph{\prd}
  {\bfseries 107} (2023) 103017}
  [\href{https://arxiv.org/abs/2211.13760}{{\ttfamily arXiv:2211.13760}}].

\bibitem{Manzari:2023gkt}
C.~A.~{Manzari}, J.~M.~{Camalich}, J.~{Spinner} and R.~{Ziegler},
  \emph{{Supernova limits on muonic dark forces}},
  \href{https://doi.org/10.1103/PhysRevD.108.103020}{\emph{\prd} {\bfseries
  108} (2023) 103020} 
  [\href{https://arxiv.org/abs/2307.03143}{{\ttfamily arXiv:2307.03143}}].

\bibitem{Bollig:2020xdr}
R.~Bollig, W.~DeRocco, P.~W.~Graham and H.-Th.~Janka, \emph{{Muons in supernovae:
  implications for the axion-muon coupling}},
  \href{https://doi.org/10.1103/PhysRevLett.125.051104}{\emph{Phys. Rev. Lett.}
  {\bfseries 125} (2020) 051104}
  [\href{https://arxiv.org/abs/2005.07141}{{\ttfamily arXiv:2005.07141}}].

\bibitem{Calibbi:2020jvd}
L.~Calibbi, D.~Redigolo, R.~Ziegler and J.~Zupan, \emph{{Looking forward to
  lepton-flavor-violating ALPs}},
  \href{https://doi.org/10.1007/JHEP09(2021)173}{\emph{JHEP} {\bfseries 09}
  (2021) 173} 
  [\href{https://arxiv.org/abs/2006.04795}{{\ttfamily arXiv:2006.04795}}].

\bibitem{Croon:2020lrf}
D.~Croon, G.~Elor, R.~K.~Leane and S.~D.~McDermott, \emph{{Supernova Muons: New
  Constraints on $Z$' Bosons, Axions and ALPs}},
  \href{https://doi.org/10.1007/JHEP01(2021)107}{\emph{JHEP} {\bfseries 01}
  (2021) 107} 
  [\href{https://arxiv.org/abs/2006.13942}{{\ttfamily arXiv:2006.13942}}].

\bibitem{Caputo:2021rux}
A.~{Caputo}, G.G.~{Raffelt} and E.~{Vitagliano}, \emph{{Muonic boson limits:
  Supernova redux}},
  \href{https://doi.org/10.1103/PhysRevD.105.035022}{\emph{Phys. Rev. D}
  {\bfseries 105} (2022) 035022}
  [\href{https://arxiv.org/abs/2109.03244}{{\ttfamily arXiv:2109.03244}}].

\bibitem{Turner:1991ax}
M.~S.~Turner, \emph{{Dirac neutrinos and SN1987A}},
  \href{https://doi.org/10.1103/PhysRevD.45.1066}{\emph{Phys. Rev. D}
  {\bfseries 45} (1992) 1066}.

\bibitem{Raffelt:1993ix}
G.~G.~{Raffelt} and D.~{Seckel}, \emph{{A selfconsistent approach to neutral
  current processes in supernova cores}},
  \href{https://doi.org/10.1103/PhysRevD.52.1780}{\emph{Phys. Rev. D}
  {\bfseries 52} (1995) 1780}
  [\href{https://arxiv.org/abs/astro-ph/9312019}{{\ttfamily arXiv:9312019}}].

\bibitem{Keil:1996ju}
W.~Keil, H.-Th.~Janka, D.~N.~Schramm, G.~Sigl, M.~S.~Turner and J.~R.~Ellis,
  \emph{{A Fresh look at axions and SN-1987A}},
  \href{https://doi.org/10.1103/PhysRevD.56.2419}{\emph{Phys. Rev. D}
  {\bfseries 56} (1997) 2419}
  [\href{https://arxiv.org/abs/astro-ph/9612222}{{\ttfamily arXiv:astro-ph/9612222}}].

\bibitem{Shin:2022ulh}
C.~S.~{Shin} and S.~{Yun}, \emph{{Dark gauge boson emission from supernova
  pions}}, \href{https://doi.org/10.1103/PhysRevD.108.055014}{\emph{\prd}
  {\bfseries 108} (2023) 055014}
  [\href{https://arxiv.org/abs/2211.15677}{{\ttfamily arXiv:2211.15677}}].

\bibitem{MartinCamalich:2020dfe}
J.~M.~{Camalich}, M.~{Pospelov}, P.~N.~H.~{Vuong}, R.~{Ziegler} and J.~{Zupan},
  \emph{{Quark Flavor Phenomenology of the QCD Axion}},
  \href{https://doi.org/10.1103/PhysRevD.102.015023}{\emph{Phys. Rev. D}
  {\bfseries 102} (2020) 015023}
  [\href{https://arxiv.org/abs/2002.04623}{{\ttfamily arXiv:2002.04623}}].

\bibitem{Camalich:2020wac}
J.~M.~{Camalich}, J.~{Terol-Calvo}, L.~{Tolos} and R.~{Ziegler},
  \emph{{Supernova Constraints on Dark Flavored Sectors}},
  \href{https://doi.org/10.1103/PhysRevD.103.L121301}{\emph{Phys. Rev. D}
  {\bfseries 103} (2021) L121301}
  [\href{https://arxiv.org/abs/2012.11632}{{\ttfamily arXiv:2012.11632}}].

\bibitem{Alonso-Alvarez:2021oaj}
G.~{Alonso-\'Alvarez}, G.~{Elor}, M.~{Escudero}, B.~{Fornal}, B.~{Grinstein}
  and J.M.~{Camalich}, \emph{{Strange physics of dark baryons}},
  \href{https://doi.org/10.1103/PhysRevD.105.115005}{\emph{Phys. Rev. D}
  {\bfseries 105} (2022) 115005}
  [\href{https://arxiv.org/abs/2111.12712}{{\ttfamily arXiv:2111.12712}}].

\bibitem{Cavan-Piton:2024ayu}
M.~{Cavan-Piton}, D.~{Guadagnoli}, M.~{Oertel}, H.~{Seong} and L.~{Vittorio},
  \emph{{Axion emission from strange matter in core-collapse SNe}},
  \href{https://arxiv.org/abs/2401.10979}{{\ttfamily arXiv:2401.10979}}.

\bibitem{Kamenik:2011vy}
J.~F.~Kamenik and C.~Smith, \emph{{FCNC portals to the dark sector}},
  \href{https://doi.org/10.1007/JHEP03(2012)090}{\emph{JHEP} {\bfseries 03}
  (2012) 090} [\href{https://arxiv.org/abs/1111.6402}{{\ttfamily arXiv:1111.6402}}].

\bibitem{Goudzovski:2022vbt}
E.~Goudzovski et~al., 
\emph{{New physics searches at kaon and hyperon factories}}, 
\href{https://doi.org/10.1088/1361-6633/ac9cee}
{\emph{Rept.\ Prog.\ Phys.} {\bfseries 86} (2023) 016201}
[\href{https://arxiv.org/abs/2201.07805}{{\ttfamily arXiv:2201.07805}}].

\bibitem{Davidson:1981zd}
A.~Davidson and K.~C.~Wali,
\emph{{Minimal flavor unifications via multigenerational Pecci-Quinn symmetry}},
\href{https://doi.org/doi:10.1103/PhysRevLett.48.11}
{\emph{\prl} {\bfseries 48} (1982), 11}

\bibitem{Wilczek:1982rv}
F.~Wilczek, \emph{{Axions and Family Symmetry Breaking}},
  \href{https://doi.org/10.1103/PhysRevLett.49.1549}{\emph{Phys. Rev. Lett.}
  {\bfseries 49} (1982) 1549}.

\bibitem{Feng:1997tn}
J.~L.~Feng, T.~Moroi, H.~Murayama and E.~Schnapka, \emph{{Third generation
  familons, b factories, and neutrino cosmology}},
  \href{https://doi.org/10.1103/PhysRevD.57.5875}{\emph{Phys. Rev.} {\bfseries
  D57} (1998) 5875} 
  [\href{https://arxiv.org/abs/hep-ph/9709411}{{\ttfamily arXiv:9709411}}].

\bibitem{Calibbi:2016hwq}
L.~Calibbi, F.~Goertz, D.~Redigolo, R.~Ziegler and J.~Zupan, \emph{{Minimal
  axion model from flavor}},
  \href{https://doi.org/10.1103/PhysRevD.95.095009}{\emph{Phys. Rev.}
  {\bfseries D95} (2017) 095009}
  [\href{https://arxiv.org/abs/1612.08040}{{\ttfamily arXiv:1612.08040}}].

\bibitem{Ema:2016ops}
Y.~Ema, K.~Hamaguchi, T.~Moroi and K.~Nakayama, \emph{{Flaxion: a minimal
  extension to solve puzzles in the standard model}},
  \href{https://doi.org/10.1007/JHEP01(2017)096}{\emph{JHEP} {\bfseries 01}
  (2017) 096} 
  [\href{https://arxiv.org/abs/1612.05492}{{\ttfamily arXiv:1612.05492}}].

\bibitem{DiLuzio:2023ndz}
L.~Di~Luzio, A.W.M.~Guerrera, X.P.~D\'\i{}az and S.~Rigolin, \emph{{On the
  IR/UV flavour connection in non-universal axion models}},
  \href{https://doi.org/10.1007/JHEP06(2023)046}{\emph{JHEP} {\bfseries 06}
  (2023) 046} 
  [\href{https://arxiv.org/abs/2304.04643}{{\ttfamily arXiv:2304.04643}}].

\bibitem{Holdom:1985ag}
B.~Holdom, \emph{{Two U(1)'s and Epsilon Charge Shifts}},
  \href{https://doi.org/10.1016/0370-2693(86)91377-8}{\emph{Phys. Lett. B}
  {\bfseries 166} (1986) 196}.

\bibitem{Dobrescu:2004wz}
B.~A.~Dobrescu, \emph{{Massless gauge bosons other than the photon}},
\href{https://doi.org/10.1103/PhysRevLett.94.151802}
{\emph{\prl} {\bfseries 94} (2005) 151802}
[\href{https://arxiv.org/abs/hep-ph/0411004}{{\ttfamily arXiv:0411004}}].

\bibitem{Gabrielli:2016cut}
E.~Gabrielli, B.~Mele, M.~Raidal and E.~Venturini, 
\emph{{FCNC decays of standard model fermions into a dark photon}},
\href{https://doi.org/10.1103/PhysRevD.94.115013}
{\emph{\prd} {\bfseries 94} (2016) 115013}
[\href{https://arxiv.org/abs/1607.05928}{{\ttfamily arXiv:1607.05928}}].

\bibitem{Eguren:2024oov}
J.~F.~Eguren, S.~Klingel, E.~Stamou, M.~Tabet and R.~Ziegler,
\emph{{Flavor phenomenology of light dark vectors}},
\href{https://doi.org/doi:10.1007/JHEP08(2024)111}
{\emph{JHEP} {\bfseries 08} (2024) 111}
[\href{https://arxiv.org/abs/2405.00108}{{\ttfamily arXiv:2405.00108}}].

\bibitem{Elor:2018twp}
G.~Elor, M.~Escudero and A.~Nelson, 
\emph{{Baryogenesis and Dark Matter from $B$ Mesons}}, 
\href{https://doi.org/10.1103/PhysRevD.99.035031}
{\emph{Phys. Rev. D} {\bfseries 99} (2019) 035031}
[\href{https://arxiv.org/abs/1810.00880}{{\ttfamily arXiv:1810.00880}}].

\bibitem{Bringmann:2018sbs}
T.~Bringmann, J.~M.~Cline and J.~M.~Cornell, \emph{{Baryogenesis from
  neutron-dark matter oscillations}},
  \href{https://doi.org/10.1103/PhysRevD.99.035024}{\emph{Phys. Rev. D}
  {\bfseries 99} (2019) 035024}
  [\href{https://arxiv.org/abs/1810.08215}{{\ttfamily arXiv:1810.08215}}].

\bibitem{Fornal:2018eol}
B.~Fornal and B.~Grinstein, \emph{{Dark Matter Interpretation of the Neutron
  Decay Anomaly}},
  \href{https://doi.org/10.1103/PhysRevLett.120.191801}{\emph{Phys. Rev. Lett.}
  {\bfseries 120} (2018) 191801}
  [\href{https://arxiv.org/abs/1801.01124}{{\ttfamily arXiv:1801.01124}}].

\bibitem{Tolos:2016hhl}
L.~Tolos, M.~Centelles and A.~Ramos, \emph{{Equation of State for Nucleonic and
  Hyperonic Neutron Stars with Mass and Radius Constraints}},
  \href{https://doi.org/10.3847/1538-4357/834/1/3}{\emph{Astrophys. J.}
  {\bfseries 834} (2017) 3} 
  [\href{https://arxiv.org/abs/1610.00919}{{\ttfamily arXiv:1610.00919}}].

\bibitem{Tolos:2017lgv}
L.~Tolos, M.~Centelles and A.~Ramos, \emph{{The Equation of State for the
  Nucleonic and Hyperonic Core of Neutron Stars}},
  \href{https://doi.org/10.1017/pasa.2017.60}{\emph{Publ. Astron. Soc.
  Austral.} {\bfseries 34} (2017) e065}
  [\href{https://arxiv.org/abs/1708.08681}{{\ttfamily arXiv:1708.08681}}].

\bibitem{Kochankovski:2022rid}
H.~Kochankovski, A.~Ramos and L.~Tolos, \emph{{Equation~of state for hot
  hyperonic neutron star matter}},
  \href{https://doi.org/10.1093/mnras/stac2671}{\emph{Mon. Not. Roy. Astron.
  Soc.} {\bfseries 517} (2022) 507}
  [\href{https://arxiv.org/abs/2206.11266}{{\ttfamily arXiv:2206.11266}}].

\bibitem{Kochankovski2024MNRAS528}
H.~{Kochankovski}, A.~{Ramos} and L.~{Tolos}, 
\emph{{Hyperonic uncertainties in neutron stars, mergers, and supernovae}},
\href{https://doi.org/10.1093/mnras/stae231}
{\emph{\mnras} {\bfseries 528} (2024) 2629} 
[\href{https://arxiv.org/abs/2309.14879}{{\ttfamily arXiv:2309.14879}}].

\bibitem{PDG}
S.~{Navas}, et~al. {(Particle Data Group collaboration)},
\emph{{Review of particle physics}},
\href{https://doi.org/10.1103/PhysRevD.110.030001}
{\emph{\prd} {\bfseries 110} (2024) 030001} 


\bibitem{LIGOScientific:2017vwq}
{\scshape LIGO Scientific, Virgo} collaboration, \emph{{GW170817: Observation
  of Gravitational Waves from a Binary Neutron Star Inspiral}},
  \href{https://doi.org/10.1103/PhysRevLett.119.161101}{\emph{Phys. Rev. Lett.}
  {\bfseries 119} (2017) 161101}
  [\href{https://arxiv.org/abs/1710.05832}{{\ttfamily arXiv:1710.05832}}].

\bibitem{NICER_Miller2019}
M.~C.~{Miller}, F.~K.~{Lamb}, A.~J.~{Dittmann}, S.~{Bogdanov}, Z.~{Arzoumanian},
  K.~C.~{Gendreau} et~al., \emph{{PSR J0030+0451 Mass and Radius from NICER Data
  and Implications for the Properties of Neutron Star Matter}},
  \href{https://doi.org/10.3847/2041-8213/ab50c5}{\emph{\apjl} {\bfseries 887}
  (2019) L24} 
  [\href{https://arxiv.org/abs/1912.05705}{{\ttfamily arXiv:1912.05705}}].

\bibitem{NICER_Watts2019}
A.~V.~{Bilous}, A.~L.~{Watts}, A.K.~{Harding}, T.E.~{Riley}, Z.~{Arzoumanian},
  S.~{Bogdanov} et~al., \emph{{A NICER View of PSR J0030+0451: Evidence for a
  Global-scale Multipolar Magnetic Field}},
  \href{https://doi.org/10.3847/2041-8213/ab53e7}{\emph{\apjl} {\bfseries 887}
  (2019) L23} 
  [\href{https://arxiv.org/abs/1912.05704}{{\ttfamily arXiv:1912.05704}}].

\bibitem{NICER_Miller2021}
M.~C.~{Miller}, F.~K.~{Lamb}, A.~J.~{Dittmann}, S.~{Bogdanov}, Z.~{Arzoumanian},
  K.C.~{Gendreau} et~al., \emph{{The Radius of PSR J0740+6620 from NICER and
  XMM-Newton Data}},
  \href{https://doi.org/10.3847/2041-8213/ac089b}{\emph{\apjl} {\bfseries 918}
  (2021) L28} 
  [\href{https://arxiv.org/abs/2105.06979}{{\ttfamily arXiv:2105.06979}}].

\bibitem{NICER_Riley2021}
T.~E.~{Riley}, A.~L.~{Watts}, P.~S.~{Ray}, S.~{Bogdanov}, S.~{Guillot},
  S.~M.~{Morsink} et~al., \emph{{A NICER View of the Massive Pulsar PSR
  J0740+6620 Informed by Radio Timing and XMM-Newton Spectroscopy}},
  \href{https://doi.org/10.3847/2041-8213/ac0a81}{\emph{\apjl} {\bfseries 918}
  (2021) L27}
  [\href{https://arxiv.org/abs/2105.06980}{{\ttfamily arXiv:2105.06980}}].

\bibitem{Typel:2009sy}
S.~Typel, G.~R{\"o}pke, T.~Kl{\"a}hn, D.~Blaschke and H.H.~Wolter,
  \emph{{Composition and thermodynamics of nuclear matter with light
  clusters}}, \href{https://doi.org/10.1103/PhysRevC.81.015803}{\emph{Phys.
  Rev. C} {\bfseries 81} (2010) 015803}
  [\href{https://arxiv.org/abs/0908.2344}{{\ttfamily arXiv:0908.2344}}].

\bibitem{Mezzacappa93a}
A.~{Mezzacappa} and S.~W.~{Bruenn}, \emph{{Type II supernovae and Boltzmann
  neutrino transport - The infall phase}},
  \href{https://doi.org/10.1086/172394}{\emph{\apj} {\bfseries 405} (1993)
  637}.

\bibitem{Mezzacappa93b}
A.~{Mezzacappa} and S.~W.~{Bruenn}, \emph{{A numerical method for solving the
  neutrino Boltzmann equation coupled to spherically symmetric stellar core
  collapse}}, \href{https://doi.org/10.1086/172395}{\emph{\apj} {\bfseries 405}
  (1993) 669}.

\bibitem{Mezzacappa93c}
A.~{Mezzacappa} and S.~W.~{Bruenn}, \emph{{Stellar core collapse - A Boltzmann
  treatment of neutrino-electron scattering}},
  \href{https://doi.org/10.1086/172791}{\emph{\apj} {\bfseries 410} (1993)
  740}.

\bibitem{Liebendorfer04}
M.~{Liebend{\"o}rfer}, O.~E.~B.~{Messer}, A.~{Mezzacappa}, S.~W.~{Bruenn},
  C.~Y.~{Cardall} and F.-K.~{Thielemann}, \emph{{A Finite Difference
  Representation of Neutrino Radiation Hydrodynamics in Spherically Symmetric
  General Relativistic Spacetime}},
  \href{https://doi.org/10.1086/380191}{\emph{\apjs} {\bfseries 150} (2004)
  263} 
  [\href{https://arxiv.org/abs/astro-ph/0207036}{{\ttfamily arXiv:0207036}}].

\bibitem{Fischer2020PhRvC101}
T.~{Fischer}, G.~{Guo}, A.~A.~{Dzhioev}, G.~{Mart{\'\i}nez-Pinedo}, M.-R.~{Wu},
  A.~{Lohs} et~al., \emph{{Neutrino signal from proto-neutron star evolution:
  Effects of opacities from charged-current-neutrino interactions and inverse
  neutron decay}},
  \href{https://doi.org/10.1103/PhysRevC.101.025804}{\emph{\prc} {\bfseries
  101} (2020) 025804} 
  [\href{https://arxiv.org/abs/1804.10890}{{\ttfamily arXiv:1804.10890}}].

\bibitem{Guo2020PhRvD102}
G.~{Guo}, G.~{Mart{\'\i}nez-Pinedo}, A.~{Lohs} and T.~{Fischer},
  \emph{{Charged-current muonic reactions in core-collapse supernovae}},
  \href{https://doi.org/10.1103/PhysRevD.102.023037}{\emph{\prd} {\bfseries
  102} (2020) 023037} 
  [\href{https://arxiv.org/abs/2006.12051}{{\ttfamily arXiv:2006.12051}}].

\bibitem{Fischer2020PhRvD102}
T.~{Fischer}, G.~{Guo}, G.~{Mart{\'\i}nez-Pinedo}, M.~{Liebend{\"o}rfer} and
  A.~{Mezzacappa}, \emph{{Muonization of supernova matter}},
  \href{https://doi.org/10.1103/PhysRevD.102.123001}{\emph{\prd} {\bfseries
  102} (2020) 123001} 
  [\href{https://arxiv.org/abs/2008.13628}{{\ttfamily arXiv:2008.13628}}].

\bibitem{LSEOS}
J.~M.~{Lattimer} and F.~{Swesty}, \emph{{A generalized equation of state for
  hot, dense matter}},
  \href{https://doi.org/10.1016/0375-9474(91)90452-C}{\emph{Nuclear Physics A}
  {\bfseries 535} (1991) 331}.

\bibitem{Hempel12}
M.~{Hempel}, T.~{Fischer}, J.~{Schaffner-Bielich} and M.~{Liebend{\"o}rfer},
  \emph{{New Equations of State in Simulations of Core-collapse Supernovae}},
  \href{https://doi.org/10.1088/0004-637X/748/1/70}{\emph{\apj} {\bfseries 748}
  (2012) 70}
  [\href{https://arxiv.org/abs/1108.0848}{{\ttfamily arXiv:1108.0848}}].

\bibitem{Hempel2010NuPhA837}
M.~{Hempel} and J.~{Schaffner-Bielich}, \emph{{A statistical model for a
  complete supernova equation of state}},
  \href{https://doi.org/10.1016/j.nuclphysa.2010.02.010}{\emph{\nphysa}
  {\bfseries 837} (2010) 210}
  [\href{https://arxiv.org/abs/0911.4073}{{\ttfamily arXiv:0911.4073}}].

\bibitem{Fischer2016EPJA52}
T.~{Fischer}, \emph{{Constraining the supersaturation density equation of state
  from core-collapse supernova simulations?. Excluded volume extension of the
  baryons}}, \href{https://doi.org/10.1140/epja/i2016-16054-9}{\emph{European
  Physical Journal A} {\bfseries 52} (2016) 54}
  [\href{https://arxiv.org/abs/1604.01629}{{\ttfamily arXiv:1604.01629}}].

\bibitem{Timmes1999}
F.~X.~{Timmes} and D.~{Arnett}, \emph{{The Accuracy, Consistency, and Speed of
  Five Equations of State for Stellar Hydrodynamics}},
  \href{https://doi.org/10.1086/313271}{\emph{\apjs} {\bfseries 125} (1999) 277}.

\bibitem{Woosley:2002zz}
S.~E.~{Woosley}, A.~{Heger} and T.~A.~{Weaver}, \emph{{The evolution and
  explosion of massive stars}},
  \href{https://doi.org/10.1103/RevModPhys.74.1015}
  {\emph{Rev. Mod. Phys.} {\bfseries 74} (2002) 1015}.

\bibitem{Jost2024arXiv240714319J}
F.~P.~{Jost}, M.~{Molero}, G.~{Nav{\'o}}, A.~{Arcones}, M.~{Obergaulinger} and
  F.~{Matteucci}, \emph{{Neutrino-driven Core-collapse Supernova Yields in
  Galactic Chemical Evolution}},
  \href{https://doi.org/10.48550/arXiv.2407.14319}{\emph{arXiv e-prints} (2024)
  arXiv:2407.14319} 
  [\href{https://arxiv.org/abs/2407.14319}{{\ttfamily arXiv:2407.14319}}].

\bibitem{Fischer2016PhRvD94_axions}
T.~{Fischer}, S.~{Chakraborty}, M.~{Giannotti}, A.~{Mirizzi}, A.~{Payez} and
  A.~{Ringwald}, \emph{{Probing axions with the neutrino signal from the next
  Galactic supernova}},
  \href{https://doi.org/10.1103/PhysRevD.94.085012}{\emph{\prd} {\bfseries 94}
  (2016) 085012} 
  [\href{https://arxiv.org/abs/1605.08780}{{\ttfamily arXiv:1605.08780}}].

\bibitem{Hempel2014ApJS214}
S.~{Banik}, M.~{Hempel} and D.~{Bandyopadhyay}, \emph{{New Hyperon Equations of
  State for Supernovae and Neutron Stars in Density-dependent Hadron Field
  Theory}}, \href{https://doi.org/10.1088/0067-0049/214/2/22}{\emph{\apjs}
  {\bfseries 214} (2014) 22} 
  [\href{https://arxiv.org/abs/1404.6173}{{\ttfamily arXiv:1404.6173}}].

\bibitem{Fiorillo:2024PhRvD108} 
D.~F.~G.~{Fiorillo}, M~{Heinlein}, H.-Th.~{Janka}, G.~{Raffelt}, E.~{Vitagliano}, R.~{Bollig},
\emph{{Supernova simulations confront SN 1987A neutrinos}},
\href{https://doi:10.1103/PhysRevD.108.083040}
{\emph{\prd} {\bfseries 108} (2023) 083040}
[\href{https://arxiv.org/abs/2308.01403}{{\ttfamily arXiv:2308.01403}}].

\bibitem{Li2024PhRvD109} 
S.~W.~{Li}, J.~F.~{Beacom}, L.~F.~{Roberts},F.~{Capozzi}
\emph{{Old data, new forensics: The first second of SN 1987A neutrino emission}},
\href{https://doi:10.1103/PhysRevD.109.083025}
{\emph{\prd} {\bfseries 109} (2024) 083025}
[\href{https://arxiv.org/abs/2306.08024}{{\ttfamily arXiv:2306.08024}}].



\end{thebibliography}

\providecommand{\href}[2]{#2}\begingroup\raggedright\endgroup

\end{document}